\documentclass[leqno,thmsa]{article}
\usepackage{amssymb,a4wide,amsmath,amsthm,epsfig,graphicx,latexsym,amsfonts,amscd}

\renewcommand{\Pr}{{\mathbb{Q}}}
\newtheorem{theorem}{Theorem}

\newtheorem{remark}[theorem]{Remark}

\makeatletter
\input{tcilatex}
\begin{document}

\title{Dangers of Bilateral Counterparty Risk: the fundamental impact of
closeout conventions}
\author{Damiano Brigo \\
{\small Dept. of Mathematics} \\
{\small King's College, London} \and Massimo Morini\thanks{%
Corresponding author. This paper expresses the views of its authors and does
not represent the institutions where the authors are working or have worked
in the past. Such institutions, including Banca IMI, are not responsible for
any use which may be made of this paper contents. We thank Giorgio
Facchinetti, Marco Bianchetti, Luigi Cefis, Martin Baxter, Andrea Bugin,
Vladimir Chorny, Josh Danziger, Igor Smirnov and other participants to the
ICBI 2010 Global Derivatives and Risk Management Conference for helpful
discussion. The remaining errors are our own. The authors would also like to
give special thanks to Andrea Pallavicini and Andrea Prampolini for
thoroughly and deeply discussing the research issues considered in this
paper.} \\
{\small Banca IMI, Intesa-SanPaolo}\\
{\small and Bocconi University, Milan}\ \ \ \ }
\date{First version March 13, 2010.\emph{\medskip \medskip\ }This version %
\today.}
\maketitle

\begin{abstract}
We analyze the practical consequences of the bilateral counterparty risk
adjustment. We point out that past literature assumes that, at the moment of
the first default, a risk-free closeout amount will be used. We argue that
the legal (ISDA) documentation suggests in many points that a substitution
closeout should be used. This would take into account the risk of default of
the survived party. We show how the bilateral counterparty risk adjustment
changes strongly when a substitution closeout amount is considered. We model
the two extreme cases of default independence and co-monotonicity, which
highlight pros and cons of both risk free and substitution closeout
formulations, and allow us to interpret the outcomes as dramatic
consequences on default contagion. Finally, we analyze the situation when
collateral is present.
\end{abstract}


\medskip

\noindent\textbf{AMS Classification Codes}: 62H20, 91B70 \newline

\noindent \textbf{keywords}: Bilateral Counterparty Risk, Credit Valuation
Adjustment, Debit Valuation Adjustment, Closeout, Default Contagion, Bond
Pricing, Default Correlation, Co-monotonic Defaults, Collateral Modeling.

\pagestyle{myheadings} \markboth{}{{\footnotesize  D. Brigo and M. Morini.
Impact of Closeout Conventions on Bilateral Counterparty Risk}}

\section{Introduction}

In this paper we analyze the practical consequences of the bilateral
counterparty risk adjustment. We point out that past literature assumes
that, at the moment of default, a risk-free closeout amount will be used.
The closeout amount is the net present value of the residual deal which is
computed when one party defaults, and that is used for default settlement. A
risk-free closeout amount is a net present value that assumes that the
surviving counterparty is default-free.

We argue here that the legal (ISDA) documentation on the settlement of a
default does not confirm this assumption. Documentation suggests in many
points that a `substitution closeout' should be used, namely a net present
value of the residual deal as computed from a third market player that is
eager to become the counterparty of the survived party for the residual
deal, replacing the defaulted one. Such a substitution closeout can be
different from a risk-free one. As a first difference, it would take into
account the risk of default of the survived party, since a real market
counterparty would never neglect it. The counterparty risk adjustments
changes strongly when a substitution closeout amount is considered, and we
show how to compute it. Also the effects at the moment of default of a
company are very different under the two closeout conventions, with some
dramatic consequences on default contagion. We show this in the following.
In the current uncertainty about which closeout will be used in future
defaults, our results should be considered carefully by the financial
community, lest there are unpleasant surprises at the moment of default of a
counterparty.

\subsubsection*{Risk-free vs substitution closeout: Practical consequences}

We analyze the practical consequences of the two different ways of computing
the closeout amount. We show on a simple derivative that the standard
formula for bilateral counterparty risk adjustment, assuming risk-free
closeout, is not consistent with the market practice on simple
uncollateralized defaultable claims such as bonds or loans or options. In
fact it causes the initial price of a deal to depend crucially on the risk
of default of one party that has no future obligations in the deal. In case
of a simple `bond' or `option' deal, it would make the price to depend on
the risk of default of the `lender', namely the party which has not future
payments to make in the deal, such as the bond buyer or the option buyer.
This feature is not consistent with current market practices and is
counterintuitive. We show that this feature can be avoided by using the
formula with substitution closeout that we introduce in this paper. In this
sense, the substitution closeout formula that we introduce appears as the
right choice to achieve consistency with market practice for
uncollateralized deals.

However, the substitution closeout formula is not immune to problems. The
main one regards the consistency with collateral computations for
collateralized deals. The collateral is calculated computing a risk-free net
present value. If the closeout is not risk-free how could the collateral
amount and the closeout amount match at default in such a way that no losses
are suffered by any party, as one would expect for collateralized deals? In
the following we propose an explanation of how this can be the case even for
a substitution closeout, but the issue would appear to be trivially solved
under a risk-free closeout.

We also observe that, if we consider the above simple uncollateralized
'bond' deals under scenarios of perfect default dependency between the two
counterparties, the risk-free closeout CVA is either the same as the
substitution closeout CVA (this happens when the 'lender' has lower credit
spreads, so that he never defaults first and we are in practice in a
unilateral CVA) or it has a behaviour that could be considered not to be
illogic, contrary to the substitution closeout. This second case happens
when the 'lender' has higher credit spreads; in this case the risk-free
closeout says that any bond should be treated as risk-free; this could be
justified by the fact that the lender always defaults first so he will never
be impacted by the default of the borrower. The situation changes when we
consider counterparties with independent default risk, as we detail below.

\subsubsection*{Forms of contagion implied by the two closeouts in different
situations}

Another important point in comparing the two closeout formulations is
understanding what happens upon default of one counterparty. In particular
we focus on the case that shows the most striking differences with the
unilateral case: the default of the counterparty with no future payments in
the deal, the `lender' in a synthetic loan transaction. Let us start with a
quite relevant problem of the risk free closeout, having possibly
destabilizing consequences.

When one of our counterparties defaults, we not only have losses if we are
creditors of the defaulted company, we may have potentially large losses
even if we are debtors, and these losses will be higher the higher our
credit spreads. This fact is so central that it is worth repeating it: if we
are debtors of a counterparty and this counterparty defaults, the accounting
of bilateral counterparty risk given in previous literature prescribes that
the default event of our creditor suddenly increases the amount of money
that we have to pay. We illustrate this by sketching a practical example
where we consider a bank that funded itself through a derivative. The bank
finds out that default of the lender makes its liabilities grow from 579
million to 927 million just as an effect of the closeout assumption. This is
a loss of 348 million on a funding deal of 1 billion notional. As we show in
the following under the assumption of independence between lender and
borrower, a substitution closeout would avoid this sudden loss and lead to a
continuous counterparty risk mark-to-market for the borrower if the lender
defaults.

Again, however, if we assume perfect dependence, there are additional
considerations to make. We present a case where the default of the lender
causes the liabilities of the borrower to decrease from 856bn to 0 if we
consider the substitution closeout. This is a stylized representation of a
not-impossible real world scenario where the default of one entity causes
the spread of its counterparties to jump extremely high. This deflates their
liabilities towards the defaulted company if we take into account their
default probability in computing the closeout amount. This shows what is the
main criticism that can be made to the substitution closeout: at the moment
of default it protects the borrowers of the defaulted company, but in doing
this it reduces the amount received by the creditors of the defaulted
company (it reduces the recovery). Thus, the risk-free closeout increases
the number of operators subject to contagion from a default, including
parties that currently seem not to think they are exposed, and this is
certainly a negative fact. On the other hand, it spreads the default losses
on higher number of parties and reduces the classic contagion channel
affecting creditors. For the creditors, this is certainly a positive fact.

It seems to us that this issue had never been studied thoroughly before, as
confirmed by the fact that an absolute standard for the computation of the
closeout amount seems not to exist among practitioners. Our mathematical and
numerical results suggest that the issue should be considered carefully,
both by practitioners and by regulators, since it has strong economic
relevance, and in particular it can affect the contagion effect of a default
event.

\subsubsection*{Structure of the paper}

In Section 2 we present the formulas introduced by previous literature for
counterparty risk adjustments. In Section 3 we analyze these formulas to
clarify what they assume about the computation of the closeout amount, and
we propose a closeout amount different from the risk-free one, namely the
substitution closeout. Then in Sections 4 and 5 we analyze ISDA
documentation. In Section 6 we perform the quantitative analysis of the
problem. We first apply the alternative formulas to a very simple payoff,
revealing that only the substitution closeout appears consistent with market
practice for bonds and loans, while the risk-free closeout introduces even
at initial time $0$ a dependence on the risk of default of the party with no
future obligation. The pattern of this unusual dependence is detailed
numerically. Then we quantify the losses or the benefit, respectively, that
a borrower would suffer in case of default of the lender when the
substitution or risk-free closeouts are applied in the respective cases of
independence and total default dependence. We have decided to consider only
these limit cases - perfect dependence and independence - because they allow
us to make claims which are as model-independent as possible, avoiding the
use of 'black-box' copula models and associated concepts such as 'default
correlation'.

In Section 7 we show that when the substitution closeout is used the
combination of a defaultable deal and a collateral agreement may lead
naturally to a default-free deal, as it should be. Finally in the Appendix
we show a number of mathematical properties, such as symmetry, of the
formula with substitution closeout that we introduce in this work.

\section{The standard risk-free closeout bilateral formulas}

We consider two parties in a derivative transaction: $A$ (investor) and $B$
(counterparty). We call $\tau ^{X}$, $R^{X}$ and $L^{X}=1-R^{X}$,
respectively, the default time, the recovery and the loss given default of
party $X$, with $X\in \left\{ A,B\right\} $. The risk-free discount factor
is
\begin{equation*}
D\left( t,T\right) =e^{-\int\nolimits_{t}^{T}r\left( s\right) ds}\text{,}
\end{equation*}%
where $r\left( t\right) $ is the risk-free short-rate. We define $\Pi
_{A}\left( t,T\right) $ to be the discounted cashflows of the derivative
from $t$ to $T$ seen from the point of view of $A$, with $\Pi _{B}\left(
t,T\right) =-\Pi _{A}\left( t,T\right) $. The net present value of the
derivative at $t$ is%
\begin{equation*}
V_{A}^{0}\left( t\right) :=\mathbb{E}_{t}\left[ \Pi _{A}\left( t,T\right) %
\right] ,
\end{equation*}%
where $\mathbb{E}_{t}$ indicates the risk-neutral expectation based on
market information up to time $t$. Notice that this, in general, includes
default monitoring, i.e. the filtration at time $t$ includes $\{\tau_X > t\}$%
. We denote by $\Pr_t$ the risk neutral probability measure conditional on
the same information at time $t$.

The subscript $A$ indicates that this value is seen from the point of view
of $A$, the superscript $0$ indicates that we are considering both parties
as default-free. Obviously, $V_{B}^{0}\left( t\right) =-V_{A}^{0}\left(
t\right) $.

The early literature on counterparty risk adjustment, see for example Brigo
and Masetti (2005), introduced `unilateral risk of default'. Here only the
default of counterparty $B$ is considered, while the investor $A$ is treated
as default free. Under this assumption, the adjusted net present value to $A$
is%
\begin{eqnarray}
&&V_{A}^{B}\left( t\right)
\begin{tabular}{l}
$=$%
\end{tabular}%
\mathbb{E}_{t}\left\{ 1_{\left\{ \tau ^{B}>T\right\} }\Pi _{A}\left(
t,T\right) \right\} +  \label{formula_unilateral_counterparty} \\
&&+\mathbb{E}_{t}\left\{ 1_{\left\{ \tau ^{B}\leq T\right\} }\left[ \Pi
_{A}\left( t,\tau ^{B}\right) +D\left( t,\tau ^{B}\right) \left( R^{B}\left(
V_{A}^{0}\left( \tau ^{B}\right) \right) ^{+}-\left( -V_{A}^{0}\left( \tau
^{B}\right) \right) ^{+}\right) \right] \right\}  \notag \\
&=&V_{A}^{0}\left( t\right) -\mathbb{E}_{t}\left[ L^{B}1_{\left\{ \tau
^{B}\leq T\right\} }D\left( t,\tau ^{B}\right) \left( V_{A}^{0}\left( \tau
^{B}\right) \right) ^{+}\right] =:V_{A}^{0}\left( t\right) -\mbox{CVA}%
_{A}\left( t\right) .  \notag
\end{eqnarray}%
The superscript $B$ indicates that this value allows for the risk of default
of $B$. Notice that we always assume both parties to be alive at $t$. The
approach is easily extended to the case when $B$ is treated as default-free,
but the default of investor $A$ is taken into account. Now the adjusted net
present value to $A$ is%
\begin{eqnarray}
&&V_{A}^{A}\left( t\right)
\begin{tabular}{l}
$=$%
\end{tabular}%
\mathbb{E}_{t}\left\{ 1_{\left\{ \tau ^{A}>T\right\} }\Pi _{A}\left(
t,T\right) \right\} +  \label{formula_unilateral_counterparty2} \\
&&+\mathbb{E}_{t}\left\{ 1_{\left\{ \tau ^{A}\leq T\right\} }\left[ \Pi
_{A}\left( t,\tau ^{A}\right) +D\left( t,\tau ^{A}\right) \left( \left(
V_{A}^{0}\left( \tau ^{A}\right) \right) ^{+}-R^{A}\left( -V_{A}^{0}\left(
\tau ^{A}\right) \right) ^{+}\right) \right] \right\}  \notag \\
&=&V_{A}^{0}\left( t\right) +\mathbb{E}_{t}\left[ L^{A}1_{\left\{ \tau
^{A}\leq T\right\} }D\left( t,\tau ^{A}\right) \cdot \left( -V_{A}^{0}\left(
\tau ^{A}\right) \right) ^{+}\right] =:V_{A}^{0}\left( t\right) +\mbox{DVA}%
_{A}\left( t\right) .  \notag
\end{eqnarray}%
The extension to the most realistic case when both $A$ and $B$ can default
is less trivial. This is called 'bilateral risk of default' and it is
introduced for interest rate swaps in Bielecki and Rutkowski (2001), Picoult
(2005) (where a simplified and approximated use of the indicators is
adopted), Gregory (2009), Brigo and Capponi (2008), and Brigo Pallavicini
and Papatheodorou (2009). In these previous works the net present value
adjusted by the default probabilities of both parties is given by%
\begin{eqnarray}
&&V_{A}\left( t\right)
\begin{tabular}{l}
$=$%
\end{tabular}%
\mathbb{E}_{t}\left\{ 1_{0}\Pi _{A}\left( t,T\right) \right\}
\label{formula_biilateral_counterparty_long} \\
&&+\mathbb{E}_{t}\left\{ 1_{A}\left[ \Pi _{A}\left( t,\tau ^{A}\right)
+D\left( t,\tau ^{A}\right) \left( \left( V_{A}^{0}\left( \tau ^{A}\right)
\right) ^{+}-R^{A}\left( -V_{A}^{0}\left( \tau ^{A}\right) \right)
^{+}\right) \right] \right\}  \notag \\
&&+\mathbb{E}_{t}\left\{ 1_{B}\left[ \Pi _{A}\left( t,\tau ^{B}\right)
+D\left( t,\tau ^{B}\right) \left( R^{B}\left( V_{A}^{0}\left( \tau
^{B}\right) \right) ^{+}-\left( -V_{A}^{0}\left( \tau ^{B}\right) \right)
^{+}\right) \right] \right\} ,  \notag
\end{eqnarray}%
where we use the following event indicators%
\begin{eqnarray*}
1_{0} &=&1_{\left\{ T<\min \left( \tau ^{A},\tau ^{B}\right) \right\} } \\
1_{A} &=&1_{\left\{ \tau ^{A}\leq \min \left( T,\tau ^{B}\right) \right\} }
\\
1_{B} &=&1-1_{A}-1_{0}=1_{\left\{ \tau ^{B}<\tau ^{A}\right\} }1_{\left\{
\tau ^{B}\leq T\right\} .}
\end{eqnarray*}%
Notice that
\begin{equation*}
V_{B}\left( t\right) =-V_{A}\left( t\right) \text{,}
\end{equation*}%
thus this formula enjoys the symmetry property that one would expect. In the
next section we analyze these formulas to understand what they implicitly
assume about the settlement of a default event, and discuss the realism of
such assumptions, and their practical consequences.

\section{The closeout problem: how much will be paid in case of a default,
really?}

Knowing which default happens first is of fundamental importance for
determining the actual payout. Thus it is convenient to rewrite formula (\ref%
{formula_biilateral_counterparty_long}) to make the order of the default
events explicit.

We define $1$ to be the first entity to default, and $2$ to be the second
one so that%
\begin{equation}
\tau ^{1}=\min \left( \tau ^{A},\tau ^{B}\right) ,\ \ \tau ^{2}=\max \left(
\tau ^{A},\tau ^{B}\right) .  \label{formula_firstseconddefaultdefinition}
\end{equation}%
With these definitions the pricing formula (\ref%
{formula_biilateral_counterparty_long}) simplifies to%
\begin{eqnarray}
V_{A}\left( t\right) &=&\mathbb{E}_{t}\left\{ 1_{0}\Pi _{A}\left( t,T\right)
\right\}  \label{formula_biilateral_counterparty_+-} \\
&&+\mathbb{E}_{t}\left\{ \left( 1_{B}-1_{A}\right) \left[ \Pi _{2}\left(
t,\tau ^{1}\right) +D\left( t,\tau ^{1}\right) \left( R^{1}\left(
V_{2}^{0}\left( \tau ^{1}\right) \right) ^{+}-\left( -V_{2}^{0}\left( \tau
^{1}\right) \right) ^{+}\right) \right] \right\} .  \notag
\end{eqnarray}%
Let us compare this bilateral pricing formula with the price $%
V_{A}^{B}\left( t\right) $ given in (\ref{formula_unilateral_counterparty}).
There the distinction between $\tau ^{1}$ and $\tau ^{2}$ is meaningless,
since only the counterparty $B$ can default, so that
\begin{eqnarray*}
\tau ^{2} &=&\tau ^{A}=+\infty , \\
\tau ^{1} &=&\tau ^{B}=:\tau
\end{eqnarray*}%
Thus,
\begin{eqnarray}
V_{A}^{B}\left( t\right) &=&\mathbb{E}_{t}\left\{ 1_{\left\{ \tau >T\right\}
}\Pi _{A}\left( t,T\right) \right\}  \label{formula_unilateral_forall} \\
&&+\mathbb{E}_{t}\left\{ 1_{\left\{ \tau \leq T\right\} }\left[ \Pi
_{A}\left( t,\tau \right) +D\left( t,\tau \right) \left( R^{1}\left(
V_{A}^{0}\left( \tau \right) \right) ^{+}-\left( -V_{A}^{0}\left( \tau
\right) \right) ^{+}\right) \right] \right\} .  \notag
\end{eqnarray}%
Now we introduce the main theme of this paper. All formulas for counterparty
risk adjustment are based on precise assumptions on what happens when there
is a default event. Let us start from the unilateral case (\ref%
{formula_unilateral_forall}). When the default of the counterparty happens
before maturity (look at the part of the formula following the indicator of
event $\left\{ \tau \leq T\right\} $), the total payout is made of two
parts: the cashflows received before default, $\Pi _{A}\left( t,\tau \right)
$, and the present value of the payout at default time $\tau $. At $\tau $
the residual deal is marked-to-market. The mark-to-market of the residual
deal at an early termination time is called \emph{closeout amount }in the
jargon of ISDA documentation. Here it is given by%
\begin{equation*}
V_{A}^{0}\left( \tau \right) =-V_{B}^{0}\left( \tau \right) ,
\end{equation*}%
where the left-hand side is seen from the perspective of the investor and
the right-hand side from the one of the counterparty. If the closeout amount
is positive to $B$, which is the defaulting party, and negative to $A$,
which has not defaulted, $A$ will pay this amount entirely to the
liquidators of the counterparty. If the closeout amount is instead positive
to $A$ and negative to $B$, the liquidators of the latter will pay to $A$
only a recovery fraction of the closeout amount. These provisions lead to
the payout at default of (\ref{formula_unilateral_forall}), given by%
\begin{equation}
\left( R^{1}\left( V_{A}^{0}\left( \tau \right) \right) -\left(
-V_{A}^{0}\left( \tau \right) \right) ^{+}\right) .
\label{formula_unilateral_defaultpayoff}
\end{equation}%
Notice that, as indicated by the superscript $0$, here the closeout amount $%
V_{A}^{0}\left( \tau \right) $ is computed treating the residual deal as a
\emph{default-free} deal. The reason for that is obvious. There are two
parties $A$ and $B$, and party $A$ is supposed default-free so it will never
default, while party $B$ has already defaulted and this is taken into
account by the fact that, in case $V_{B}^{0}\left( \tau \right) <0$, $B$
will pay only a recovery fraction of the default-free closeout amount. This
default-free closeout is also a substitution closeout, in the sense that, if
$A$ wanted to substitute the defaulted deal with another one where the
counterparty is default-free, the counterparty would ask $A$ to pay $%
V_{A}^{0}\left( \tau \right) $, a risk-free closeout since both parties are
risk-free.

Now let us look at the pricing formula (\ref%
{formula_biilateral_counterparty_+-}) for bilateral risk of default. The
payout at default is now given by%
\begin{equation*}
\left( 1_{B}-1_{A}\right) \left[ R^{1}\left( V_{2}^{0}\left( \tau
^{1}\right) \right) ^{+}-\left( -V_{2}^{0}\left( \tau ^{1}\right) \right)
^{+}\right]
\end{equation*}%
Here both $A$ and $B$ can default, and what matters is who defaults first.
If the counterparty $B$ defaults first, then $1_{B}=1$, $1_{A}=0$, $\tau
^{1}=\tau ^{B}$ and we have the same payout as in the unilateral case of (%
\ref{formula_unilateral_defaultpayoff}). If the investor $A$ defaults first,
the payout is reversed: when the closeout amount is positive to the
defaulted investor, this amount is received fully, while if it is negative
only a recovery fraction will be paid to the counterparty, leading to%
\begin{equation*}
\left[ \left( V_{A}^{0}\left( \tau ^{1}\right) \right) ^{+}-R^{1}\left(
-V_{A}^{0}\left( \tau ^{1}\right) \right) ^{+}\right] .
\end{equation*}%
Notice that, like in the unilateral case, also with bilateral risk of
default the closeout amount \emph{is computed treating the residual deal as
default-free}
\begin{equation}
V_{1}^{0}\left( \tau ^{1}\right) =-V_{2}^{0}\left( \tau ^{1}\right) .
\label{formula_bilateralriskfreecloseout}
\end{equation}%
We are again excluding the possibility of default of either party. Is this
assumption as obviously justified here as it was in the unilateral case? Not
quite. Only the assumption of ignoring the risk of default of $1$ is
justified obviously. In fact $1$ has defaulted, and this is accounted for
correctly by computing a closeout amount where there is no possibility of
another default of $1$, and then, in case this amount is negative to $1$,
assuming that $1$ will pay only a recovery fraction of it. But the other
party $2$ has not defaulted, and now it is not true that it will never
default in the future. There is a non-negligible probability that it
defaults before the maturity $T$ of the residual deal. Thus it is unclear
why the mark-to-market of a residual deal where one of the two parties has
not yet defaulted and can default in the future should be treated as default
free. Here the \emph{risk-free closeout amount} (\ref%
{formula_bilateralriskfreecloseout}) is not a substitution closeout. In fact
if the survived party $2$ wanted to substitute the defaulted deal with
another one where the market counterparty is default free, the counterparty
would ask $2$ to pay not the opposite of (\ref%
{formula_bilateralriskfreecloseout}) but%
\begin{equation}
V_{2}^{2}\left( \tau ^{1}\right) =-V_{1}^{2}\left( \tau ^{1}\right) ,
\label{formula_bilateralSubstitutioncloseout}
\end{equation}%
because the market counterparty cannot ignore the default risk of party $2$
from $\tau ^{1}$ to the maturity $T$ of the residual deal. Thus the amount (%
\ref{formula_bilateralSubstitutioncloseout}) will be called in the following
\emph{substitution closeout amount}. It is given by (\ref%
{formula_unilateral_counterparty}) when $1=B$ and by (\ref%
{formula_unilateral_counterparty2}) when $1=A$.

Allowing for this computation of the closeout amount, the pricing formula in
case of bilateral counterparty risk becomes
\begin{eqnarray}
&&\hat{V}_{A}\left( t\right)
\begin{tabular}{l}
$=$%
\end{tabular}%
\mathbb{E}_{t}\left\{ 1_{0}\Pi _{A}\left( t,T\right) \right\}
\label{formula_bilateral righr} \\
&&+\mathbb{E}_{t}\left\{ 1_{A}\left[ \Pi _{A}\left( t,\tau ^{A}\right)
+D\left( t,\tau ^{A}\right) \left( \left( V_{A}^{B}\left( \tau ^{A}\right)
\right) ^{+}-R^{A}\left( -V_{A}^{B}\left( \tau ^{A}\right) \right)
^{+}\right) \right] \right\}  \notag \\
&&+\mathbb{E}_{t}\left\{ 1_{B}\left[ \Pi _{A}\left( t,\tau ^{B}\right)
+D\left( t,\tau ^{B}\right) \left( R^{B}\left( V_{A}^{A}\left( \tau
^{B}\right) \right) ^{+}-\left( -V_{A}^{A}\left( \tau ^{B}\right) \right)
^{+}\right) \right] \right\} .  \notag
\end{eqnarray}%
In the rest of the paper we analyze this formula under two points of view.
First, we want to understand if it is more appropriate than formula (\ref%
{formula_biilateral_counterparty_long}) used in the previous literature. For
this analysis we consider

1) the ISDA documentation on derivatives, to understand the legal
prescriptions and the financial rationale that should be followed in
computing the closeout amount

2) which one between the substitution closeout amount assumed by (\ref%
{formula_bilateral righr}) and the risk-free one assumed by (\ref%
{formula_biilateral_counterparty_long}) is simpler to be applied in case of
a real default.

3) the financial effects of using (\ref{formula_bilateral righr}) rather
than (\ref{formula_biilateral_counterparty_long}), to see which one
minimizes the default contagion in case of default, which one is more
consistent with market standards on consolidated financial products and
which one fits better with market practices to minimize default risk such as
CSA collateral agreements.

One observation is in order about this plan of analysis . The reader may
think that the first step to take is to ask experienced practitioners which
approach is applied in practice for the computation of closeout. We have
done this (and we thank for that in particular the participants to Global
Derivatives 2010 in Paris) and surprisingly enough we have found different
opinions. This variety of opinions can be related to the fact that the ISDA
documentation had given a (relatively) open definition of closeout amount,
and more importantly to the fact that full awareness of the importance of
CVA and DVA has arisen just \emph{after} Lehman's default, the last
important default that may work as a benchmark. At that time
\textquotedblleft Libor discounting\textquotedblright\ was mainly used for
closeout, and this is somehow in-between a risk-free discounting and a
discounting taking full consideration of the risk of default of the
remaining party.

Then, in the appendix, we analyze the new formula (\ref{formula_bilateral
righr}) under a mathematical point of view. We first show that it enjoys the
symmetry property $\hat{V}_{A}\left( t\right) =-\hat{V}_{B}\left( t\right) $
as well. Then we show that is equivalent to%
\begin{equation*}
\begin{tabular}{l}
$\hat{V}_{A}\left( t\right)
\begin{tabular}{l}
$=$%
\end{tabular}%
\mathbb{E}_{t}\left\{ \Pi _{A}\left( t,T\right) \right\} \medskip $ \\
$+\mathbb{E}_{t}\left\{ 1_{1}\left[ D\left( t,\tau ^{A}\right) \left( -%
\mathbb{E}_{\tau ^{A}}\left[ L^{B}1_{\left\{ \tau ^{B}\leq T\right\}
}D\left( \tau ^{A},\tau ^{B}\right) \cdot \left( V_{A}^{0}\left( \tau
^{B}\right) \right) ^{+}\right] \right. \right. \right. \medskip $ \\
$\left. \left. \left. +\left( 1-R^{A}\right) \left( -V_{A}^{0}\left( \tau
^{A}\right) +\mathbb{E}_{\tau ^{A}}\left[ L^{B}1_{\left\{ \tau ^{B}\leq
T\right\} }D\left( \tau ^{A},\tau ^{B}\right) \cdot \left( V_{A}^{0}\left(
\tau ^{B}\right) \right) ^{+}\right] \right) ^{+}\right) \right] \right\}
\medskip $ \\
$+\mathbb{E}_{t}\left\{ 1_{2}\left[ D\left( t,\tau ^{B}\right) \left(
\mathbb{E}_{\tau ^{B}}\left[ L^{A}1_{\left\{ \tau ^{A}\leq T\right\}
}D\left( \tau ^{B},\tau ^{A}\right) \cdot \left( -V_{A}^{0}\left( \tau
^{A}\right) \right) ^{+}\right] \right. \right. \right. \medskip $ \\
$\left. \left. \left. -\left( 1-R^{B}\right) \left( V_{A}^{0}\left( \tau
^{B}\right) +\mathbb{E}_{\tau ^{B}}\left[ L^{A}1_{\left\{ \tau ^{A}\leq
T\right\} }D\left( \tau ^{B},\tau ^{A}\right) \cdot \left( -V_{A}^{0}\left(
\tau ^{A}\right) \right) ^{+}\right] \right) ^{+}\right) \right] \right\} ,$%
\end{tabular}%
\end{equation*}%
and that, using the definitions in (\ref%
{formula_firstseconddefaultdefinition}) and recalling that $V_{B}^{0}\left(
t\right) =-V_{A}^{0}\left( t\right) $, it can be simplified in
\begin{eqnarray*}
&&\hat{V}_{A}\left( t\right)
\begin{tabular}{l}
$=$%
\end{tabular}%
\mathbb{E}_{t}\left\{ 1_{0}\Pi _{A}\left( t,T\right) \right\} \\
&&+\mathbb{E}_{t}\left\{ \left( 1_{B}-1_{A}\right) \left[ \Pi _{2}\left(
t,\tau ^{1}\right) +D\left( t,\tau ^{1}\right) \left( R^{1}\left(
V_{2}^{2}\left( \tau ^{1}\right) \right) ^{+}-\left( V_{1}^{2}\left( \tau
^{1}\right) \right) ^{+}\right) \right] \right\} .
\end{eqnarray*}

\section{A `legal analysis': the ISDA closeout amount protocol}

The document that should set a standard for the computation of the close-out
amount in case of a default event is the ISDA (2009) Close-out Amount
Protocol. Nowhere in this document one finds a precise formula for the
computation of the close-out amount, however one can find there practical
principles that can shed some light on the issue we are considering. We read
at page 13 that "If the Early Termination Date results from an Event of
Default", this early termination will be settled by the transfer of "the
Close-out Amount or Close-out Amounts (whether positive or negative)
determined by the Non-defaulting Party". The non-defaulting party that
determines the closeout amount is party $2$ in our notation. Then at page 15
we have the following prescription: "In determining a Close-out Amount, the
Determining Party may consider any relevant information, including, without
limitation, one or more of the following types of information: (i)
quotations (either firm or indicative) for replacement transactions supplied
by one or more third parties that may take into account the creditworthiness
of the Determining Party at the time the quotation is provided". This is in
contrast with the default-free closeout amount $V_{2}^{0}\left( \tau
^{1}\right) $ prescribed by the classic formula, and seems instead
consistent with the substitution-cost closeout amount $V_{2}^{2}\left( \tau
^{1}\right) $ prescribed by the formula given in this paper, that includes
the risk of default of the survived party $2$. The ISDA documentation is not
so strict to make this a binding prescription - the document speaks of a
determining party that \emph{may} take into account its own
creditworthiness. Thus the risk-free closeout amount is not excluded.

Various other points in the documentation confirm that a substitution
closeout is more likely. One of the clearest is in the Market Review of OTC
Derivative Bilateral Collateralization Practices, published by ISDA on March
1, 2010, that says: "Upon default close-out, valuations will in many
circumstances reflect the replacement cost of transactions calculated at the
terminating party's bid or offer side of the market, \emph{and will often
take into account the credit-worthiness of the terminating
party\textquotedblright }.

The same document, however, points out that this substitution closeout risks
being a problem for collateralized derivatives, since it seems at odds with
the computation of collateral amount: \textquotedblleft However, it should
be noted that Exposure is calculated at mid-market levels so as not to
penalize one party or the other. As a result of this, the amount of
collateral held to secure Exposure may be more or less than the termination
payment determined upon a close-out". We analyze this issue in detail in
Section \ref{section_collateral}.

\section{Risk-free vs substitution closeout: what is simpler to compute?}

The above analysis shows a financial rationale in the ISDA documents in
favor of a substitution closeout, but it also shows that the same documents
leave room for other solutions. There is one point in strong favor of a
risk-free closeout: the simplicity of computation, since it does not require
an assessment of the risk of default of the survived party. In the
settlement of a default, if a risk-free closeout is used, then all contracts
with the same payoff would have the same closeout value, irrespectively of
the counterparty, which is an important simplification. This risk-free
amount would not be difficult to compute in the market because it would
correspond to the mark-to-market of a deal equal to the residual deal but
for the fact of being collateralized. Of course some parties could find it
unfair. In the end the defaulted deal was not collateralized, and at
inception it had different prices for different parties because of the
different default risks of the parties. Some may not accept now a unique
'collateralized' closeout amount. In spite of this, the homogeneity of the
risk-free closeout amount is a strong point in favor of this solution, maybe
just as a simpler approximation to the substitution closeout, if the error
turns out to be not too big.

The above considerations do not appear sufficient to clarify our minds and
settle the matter without a more quantitative analysis. We perform such
analysis in the following. Since we are dealing with a matter that would
affect even the simplest derivative in case of default, we find an analysis
with very simple - albeit not unrealistic - models and transactions to be
much more instructive than the application of some elaborate model to a
complex derivative, that would just risk increasing the confusion. Once we
have sorted out the basic case, we can think of addressing more complex
payouts in future work.

\section{A quantitative analysis: pricing a bond payoff}

We perform a\ mathematical and numerical analysis of the consequences of the
two approaches. We consider a contract where party $B$ enters at time $0$
into the commitment to pay \ a unit of money at time $T$ to party $A$. This
claim has exactly the same payoff as the prototypical zero-coupon bond or
loan of the textbooks on finance. Party $B$ is the borrower or bond issuer,
and $A$ is the lender or bond holder. This payoff is convenient for our
purposes because there is a consolidated standard on how to price it. We
will see which one between risk-free and substitution liquidation is
consistent with the practice developed in the loan and bond markets.

The comparison with the bond market is particularly interesting for a
further reason. When bilateral counterparty risk was first introduced, there
was some discussion in the market about one consequence of it: when a bank
includes its own risk of default in the pricing of a deal, it can actually
book a profit when there is an increase of its credit spreads. This is a
rather bizarre fact. However supporters of this approach pointed out that
this already happens for banks with reference to bond issuances. Banks have
the so-called \emph{fair value option}, namely the possibility to account
for issued bonds in their balance-sheets at mark-to-market. When this is
done, the bond liabilities decrease in value when credit spreads increase,
and the bank can mark a profit. This consistency with the treatment of bonds
has contributed to make the DVA more accepted in the market. Below we will
show some practical effects of bilateral counterparty risk adjustments under
risk-free closeout or substitution closeout. By observing these effects on a
bond-like payoff, we are using the payoff that has already been the main
reference for understanding the appropriateness of counterparty risk
adjustments.

\subsection{Pricing a bond under risk-free or substitution closeout}

In the following we take deterministic interest rates.

\subsubsection{The unilateral case}

First we evaluate this 'derivative bond' deal using unilateral formulas,
namely considering only the default of the borrower counterparty $B$ (here
the default of $A$ is neglected from the payout irrespectively of the fact
that it may default or not). We apply (\ref{formula_unilateral_counterparty}%
) to this payoff, getting
\begin{eqnarray}  \label{formula_unilateral_counterparty_bis}
\hspace{1cm}V_{A}^{B}\left( t\right) &=&\mathbb{E}_{t}\left\{ 1_{\left\{
\tau ^{B}>T\right\} }e^{-\int\nolimits_{t}^{T}r\left( s\right) ds}\right\} +%
\mathbb{E}_{t}\left\{ 1_{\left\{ \tau ^{B}\leq T\right\} }\left[
e^{-\int\nolimits_{t}^{\tau ^{B}}r\left( s\right) ds}R^{B}\left(
e^{-\int\nolimits_{\tau ^{B}}^{T}r\left( s\right) ds}\right) ^{+}\right]
\right\}  \label{formula_bond under the risk or borrower} \\
&=&e^{-\int\nolimits_{t}^{T}r\left( s\right) ds}\mathbb{E}_{t}\left[
1_{\left\{ \tau ^{B}>T\right\} }\right] +R^{B}e^{-\int\nolimits_{t}^{T}r%
\left( s\right) ds}\mathbb{E}_{t}\left[ 1_{\left\{ \tau ^{B}\leq T\right\} }%
\right]  \notag
\end{eqnarray}%
This is the standard formula for the pricing of a defaultable bond or loan.
We have%
\begin{equation*}
V_{A}^{B}\left( 0\right) =e^{-\int\nolimits_{0}^{T}r\left( s\right) ds}\Pr
\left( \tau ^{B}>T\right) +R^{B}e^{-\int\nolimits_{0}^{T}r\left( s\right)
ds}\Pr \left( \tau ^{B}\leq T\right) ,
\end{equation*}%
which says that the price of a defaultable bond equals the price of a
default-free bond multiplied by the survival probability of the issuer, plus
a recovery part received when the issuer defaults.

By applying (\ref{formula_unilateral_counterparty2}) to this payoff we can
compute easily also $V_{A}^{A}\left( t\right) $, the value when only the
default of the lender $A$ is taken into account. We have%
\begin{equation}
V_{A}^{A}\left( t\right) =\mathbb{E}_{t}\left\{ 1_{\left\{ \tau
^{A}>T\right\} }e^{-\int\nolimits_{t}^{T}r\left( s\right) ds}\right\} +%
\mathbb{E}_{t}\left\{ 1_{\left\{ \tau ^{A}\leq T\right\}
}e^{-\int\nolimits_{t}^{\tau ^{A}}r\left( s\right) ds}\left(
e^{-\int\nolimits_{\tau ^{A}}^{T}r\left( s\right) ds}\right) ^{+}\right\}
=e^{-\int\nolimits_{t}^{T}r\left( s\right) ds}.
\label{formula_bond under the risk of lender}
\end{equation}%
We have obtained the price of a risk-free bond, thus the formula says that
in a loan or bond what matters is the risk of default of the borrower. If we
consider only risk of default for the lender, and not default-risk of the
borrower, we get just the price of a default-free loan or bond. We have no
influence of the risk of default of a party, the lender, that in this
contract has no future obligations. Both formula (\ref{formula_bond under
the risk or borrower}) and (\ref{formula_bond under the risk of lender}) are
in line with market practice.

\subsubsection{The bilateral case with substitution closeout}

Now we price the deal considering the default risk of both parties, \emph{%
and assuming first a substitution closeout}. We apply formula (\ref%
{formula_bilateral righr}) introduced in this paper, putting (\ref%
{formula_bond under the risk or borrower}) and (\ref{formula_bond under the
risk of lender}) into this formula. We get%
\begin{eqnarray*}
&&\hat{V}_{A}\left( t\right)
\begin{tabular}{l}
$=$%
\end{tabular}%
\mathbb{E}_{t}\left\{ 1_{0}e^{-\int\nolimits_{t}^{T}r\left( s\right)
ds}\right\} \\
&&+\mathbb{E}_{t}\left\{ 1_{A}\left[ e^{-\int\nolimits_{t}^{\tau
^{A}}r\left( s\right) ds}\left( e^{-\int\nolimits_{\tau ^{A}}^{T}r\left(
s\right) ds}\mathbb{E}_{\tau ^{A}}\left[ 1_{\left\{ \tau ^{B}>T\right\} }%
\right] +R^{B}e^{-\int\nolimits_{\tau ^{A}}^{T}r\left( s\right) ds}\mathbb{E}%
_{\tau ^{A}}\left[ 1_{\left\{ \tau ^{B}\leq T\right\} }\right] \right) ^{+}%
\right] \right\} \\
&&+\mathbb{E}_{t}\left\{ 1_{B}\left[ e^{-\int\nolimits_{t}^{\tau
^{B}}r\left( s\right) ds}R^{B}\left( e^{-\int\nolimits_{\tau
^{B}}^{T}r\left( s\right) ds}\right) ^{+}\right] \right\} \\
&=&\mathbb{E}_{t}\left\{ 1_{0}e^{-\int\nolimits_{t}^{T}r\left( s\right)
ds}\right\} \\
&&+\mathbb{E}_{t}\left\{ 1_{A}\left[ e^{-\int\nolimits_{t}^{T}r\left(
s\right) ds}\mathbb{E}_{\tau ^{A}}\left[ 1_{\left\{ \tau ^{B}>T\right\} }%
\right] +R^{B}e^{-\int\nolimits_{t}^{T}r\left( s\right) ds}\mathbb{E}_{\tau
^{A}}\left[ 1_{\left\{ \tau ^{B}\leq T\right\} }\right] \right] \right\} \\
&&+\mathbb{E}_{t}\left\{ 1_{B}\left[ R^{B}e^{-\int\nolimits_{t}^{T}r\left(
s\right) ds}\right] \right\}
\end{eqnarray*}%
Notice that
\begin{equation*}
1_{A}=1_{\left\{ \tau ^{A}\leq \min \left( T,\tau ^{B}\right) \right\} }
\end{equation*}%
is $\tau ^{A}$-measurable, namely it is known at $\tau ^{A}$, so that the
second one of the three terms of $\hat{V}_{A}\left( t\right) $ above can be
rewritten as
\begin{eqnarray*}
&&\mathbb{E}_{t}\left\{ 1_{A}\left[ e^{-\int\nolimits_{t}^{T}r\left(
s\right) ds}\mathbb{E}_{\tau ^{A}}\left[ 1_{\left\{ \tau ^{B}>T\right\} }%
\right] +R^{B}e^{-\int\nolimits_{t}^{T}r\left( s\right) ds}\mathbb{E}_{\tau
^{A}}\left[ 1_{\left\{ \tau ^{B}\leq T\right\} }\right] \right] \right\} \\
&=&e^{-\int\nolimits_{t}^{T}r\left( s\right) ds}\mathbb{E}_{t}\left[ \mathbb{%
E}_{\tau ^{A}}\left[ 1_{A}1_{\left\{ \tau ^{B}>T\right\} }\right] \right]
+R^{B}e^{-\int\nolimits_{t}^{T}r\left( s\right) ds}\mathbb{E}_{t}\left[
\mathbb{E}_{\tau ^{A}}\left[ 1_{A}1_{\left\{ \tau ^{B}\leq T\right\} }\right]
\right] \\
&=&e^{-\int\nolimits_{t}^{T}r\left( s\right) ds}\mathbb{E}_{t}\left[
1_{A}1_{\left\{ \tau ^{B}>T\right\} }\right] +R^{B}e^{-\int%
\nolimits_{t}^{T}r\left( s\right) ds}\mathbb{E}_{t}\left[ 1_{A}1_{\left\{
\tau ^{B}\leq T\right\} }\right]
\end{eqnarray*}%
where in the last passage we have used the law of iterated expectations. Now
in $\hat{V}_{A}\left( t\right) $ we can factor together the terms that are
multiplied by $R^{B}$ and those that are not, getting%
\begin{eqnarray*}
\hat{V}_{A}\left( t\right) &=&e^{-\int\nolimits_{t}^{T}r\left( s\right) ds}%
\mathbb{E}_{t}\left[ 1_{0}\right] +e^{-\int\nolimits_{t}^{T}r\left( s\right)
ds}\mathbb{E}_{t}\left[ 1_{A}1_{\left\{ \tau ^{B}>T\right\} }\right] \\
&&+R^{B}e^{-\int\nolimits_{t}^{T}r\left( s\right) ds}\mathbb{E}_{t}\left[
1_{A}1_{\left\{ \tau ^{B}\leq T\right\} }\right] +R^{B}e^{-\int%
\nolimits_{t}^{T}r\left( s\right) ds}\mathbb{E}_{t}\left[ 1_{B}\right] \\
&=&e^{-\int\nolimits_{t}^{T}r\left( s\right) ds}\mathbb{E}_{t}\left[
1_{0}+1_{A}1_{\left\{ \tau ^{B}>T\right\} }\right] \\
&&+R^{B}e^{-\int\nolimits_{t}^{T}r\left( s\right) ds}\mathbb{E}_{t}\left[
1_{A}1_{\left\{ \tau ^{B}\leq T\right\} }+1_{B}\right]
\end{eqnarray*}%
Now we concentrate on the indicators. It is easy to see that
\begin{eqnarray*}
1_{A}1_{\left\{ \tau ^{B}>T\right\} } &=&1_{\left\{ \tau ^{A}\leq \min
\left( T,\tau ^{B}\right) \right\} }1_{\left\{ \tau ^{B}>T\right\}
}=1_{\left\{ \tau ^{A}\leq T\right\} }1_{\left\{ \tau ^{B}>T\right\} }, \\
1_{A}1_{\left\{ \tau ^{B}\leq T\right\} } &=&1_{\left\{ \tau ^{A}\leq \min
\left( T,\tau ^{B}\right) \right\} }1_{\left\{ \tau ^{B}\leq T\right\}
}=1_{\left\{ \tau ^{A}\leq \tau ^{B}\right\} }1_{\left\{ \tau ^{B}\leq
T\right\} } \\
1_{0} &=&1_{\left\{ T<\min \left( \tau ^{A},\tau ^{B}\right) \right\}
}=1_{\left\{ T<\tau ^{A}\right\} }1_{\left\{ T<\tau ^{B}\right\} },
\end{eqnarray*}%
leading to
\begin{eqnarray*}
\hat{V}_{A}\left( t\right) &=&e^{-\int\nolimits_{t}^{T}r\left( s\right) ds}%
\mathbb{E}_{t}\left[ 1_{\left\{ T<\tau ^{A}\right\} }1_{\left\{ T<\tau
^{B}\right\} }+1_{\left\{ \tau ^{A}\leq T\right\} }1_{\left\{ \tau
^{B}>T\right\} }\right] \\
&&+R^{B}e^{-\int\nolimits_{t}^{T}r\left( s\right) ds}\mathbb{E}_{t}\left[
1_{\left\{ \tau ^{A}\leq \tau ^{B}\right\} }1_{\left\{ \tau ^{B}\leq
T\right\} }+1_{\left\{ \tau ^{B}<\tau ^{A}\right\} }1_{\left\{ \tau ^{B}\leq
T\right\} }\right] .
\end{eqnarray*}%
Since $1_{\left\{ T<\tau ^{A}\right\} }+1_{\left\{ \tau ^{A}\leq T\right\}
}=1_{\left\{ \tau ^{A}\leq \tau ^{B}\right\} }+1_{\left\{ \tau ^{B}<\tau
^{A}\right\} }=1$, we have%
\begin{eqnarray}
\hat{V}_{A}\left( t\right) &=&e^{-\int\nolimits_{t}^{T}r\left( s\right) ds}%
\mathbb{E}_{t}\left[ 1_{\left\{ \tau ^{B}>T\right\} }\right]
+R^{B}e^{-\int\nolimits_{t}^{T}r\left( s\right) ds}\mathbb{E}_{t}\left[
1_{\left\{ \tau ^{B}\leq T\right\} }\right]
\label{formula_bilateral righr_for bond} \\
&=&e^{-\int\nolimits_{t}^{T}r\left( s\right) ds}\left[ \Pr_{t}\left( \tau
^{B}>T\right) +R^{B}\Pr_{t}\left( \tau ^{B}\leq T\right) \right] .  \notag
\end{eqnarray}%
In spite of the somewhat lengthy computations, we have found a very simple
and reasonable result. We have $\hat{V}_{A}(t)=V_{A}^{B}(t)$, and we have
that the risk of default of the bond-holder in a bond, or the lender in a
loan, does not influence the value of the contract.

\begin{remark}
\label{filtrationdelirium}We point out that in the above formula for $\hat{V}%
_{A}\left( t\right) $ there can be dependence from the risk of default of
the lender through the terms $\Pr_{t}\left( \tau ^{B}>T\right) $ and $%
\Pr_{t}\left( \tau ^{B}\leq T\right) $ when the formula is evaluated at a
future time $t>0$. In fact in some models terms such as $\Pr_{t}\left( \tau
^{B}>T\right) $ do depend on the the risk of default of $A$. For example,
when using a copula model, at times $t>0$ the fact that a correlated
counterparty has defaulted or not by $t$ changes the default probability of
a counterparty still alive, since dependency is set on unobservable latent
variables (the exponential triggers), about which one can get information by
observing if correlated companies have defaulted or not. In other models
such as the structural first passage model of Black and Cox (1976) or the
multivariate exponential Marshall Olkin (1967) model this does not happen.
We point out that however also within models, like copulas, where terms such
as $\Pr_{t}\left( \tau ^{B}>T\right) $ do depend on the the risk of default
of $A$, pricing is always made assuming $t=0$, where even in these models
there is independence from the the risk of default of $A$. The situation is
different instead when in a copula we make a forward valuation where we need
to assume $t>0$, as we do in the following when we asses the value of a deal
just before and just after a default. There we show how conditioning on
information about $A$ modifies $\Pr_{t}\left( \tau ^{B}>T\right) $.
\end{remark}

We can conclude that, although bilateral risk of default matters in general
for bilateral contracts, as confirmed by the fact that $\hat{V}_{A}\left(
0\right) $ is in general much more complex that $V_{A}^{B}\left( 0\right) $,
with substitution closeout we have that when the contract has no future
obligations for a party $A$ the risk of default of $A$ does not influence
the price. Only the risk of default of the bond-issuer or borrower matters
for valuation. This result is also in line with market practice.\footnote{%
The reader may think that the risk of default of the lender could in
practice influence the price of a bond or loan through its effect on the
cost of funding for the lender. See Morini and Prampolini (2010) for a
discussion on this. However, when liquidity costs are note considered, like
in this paper, this effect does not exist and, like in classic bond pricing,
one expects the risk of default of the holder not to influence the price of
a bond.}

\subsubsection{The bilateral case with risk-free closeout}

Now we apply instead \emph{the formula (\ref%
{formula_biilateral_counterparty_long}) that assumes a risk-free closeout}.
Since $V_{A}^{0}\left( \tau \right) =e^{-\int\nolimits_{\tau }^{T}r\left(
s\right) ds}$, we have
\begin{eqnarray*}
&&V_{A}\left( t\right)
\begin{tabular}{l}
$=$%
\end{tabular}%
\mathbb{E}_{t}\left\{ 1_{0}e^{-\int\nolimits_{t}^{T}r\left( s\right)
ds}\right\} +\mathbb{E}_{t}\left\{ 1_{A}\left[ e^{-\int\nolimits_{t}^{\tau
^{A}}r\left( s\right) ds}\left( e^{-\int\nolimits_{\tau ^{A}}^{T}r\left(
s\right) ds}\right) ^{+}\right] \right\} \\
&&+\mathbb{E}_{t}\left\{ 1_{B}\left[ e^{-\int\nolimits_{t}^{\tau
^{B}}r\left( s\right) ds}R^{B}\left( e^{-\int\nolimits_{\tau
^{B}}^{T}r\left( s\right) ds}\right) ^{+}\right] \right\} \\
&=&e^{-\int\nolimits_{t}^{T}r\left( s\right) ds}\mathbb{E}_{t}\left[
1_{0}+1_{A}\right] +e^{-\int\nolimits_{t}^{T}r\left( s\right) ds}R^{B}%
\mathbb{E}_{t}\left[ 1_{B}\right] .
\end{eqnarray*}%
We can write%
\begin{equation}
V_{A}\left( t\right) =e^{-\int\nolimits_{t}^{T}r\left( s\right) ds}\Pr_{t}
\left[ T<\min \left( \tau ^{A},\tau ^{B}\right) \cup \tau ^{A}\leq \min
\left( T,\tau ^{B}\right) \right] +e^{-\int\nolimits_{t}^{T}r\left( s\right)
ds}R^{B}\Pr_{t}\left[ \tau ^{B}<\tau ^{A}\cap \tau ^{B}\leq T\right] .
\label{eq:VARFmain}
\end{equation}%
Playing with indicators one can also obtain the alternative expressions
\begin{eqnarray}
V_{A}(t) &=&e^{-\int\nolimits_{t}^{T}r\left( s\right) ds}\left( \Pr_{t}[\tau
^{B}>\min (\tau ^{A},T)]+R^{B}\Pr_{t}[\tau ^{B}<\min (\tau ^{A},T)]\right)
\label{eq:VARFalt} \\
&=&e^{-\int\nolimits_{t}^{T}r\left( s\right) ds}\left( \Pr_{t}[\tau
^{B}>T]+\Pr_{t}[\tau ^{A}<\tau ^{B}<T]+R^{B}\Pr_{t}[\tau ^{B}<\min (\tau
^{A},T)]\right)  \notag
\end{eqnarray}

\subsubsection{Comparing the two closeout formulations}

Reaching this result for $V_{A}$ applied to a `derivative bond' has been far
easier than reaching the analogous one for $\hat{V}_{A}$, but the result
looks more complex. In fact it introduces a dependence on the risk of
default of the lender even at time $0$, and on the exact individuation of
the first default, that was not there in $\hat{V}_{A}$. This is in contrast
with the market practice in the bond or loan markets. In particular, let us
compare $V_{A}\left( t\right) $ with $\hat{V}_{A}\left( t\right) $. Since
\begin{eqnarray*}
1_{0}+1_{A} &=&1_{\left\{ \tau ^{B}>T\right\} }1_{\left\{ \tau
^{A}>T\right\} }+1_{\left\{ \tau ^{B}\geq \tau ^{A}\right\} }1_{\left\{ \tau
^{A}\leq T\right\} } \\
&\geq & \\
1_{\left\{ \tau ^{B}>T\right\} } &=&1_{\left\{ \tau ^{B}>T\right\}
}1_{\left\{ \tau ^{A}>T\right\} }+1_{\left\{ \tau ^{B}>T\right\} }1_{\left\{
\tau ^{A}\leq T\right\} }
\end{eqnarray*}%
and $R^{B}\leq 1$, we have%
\begin{equation*}
V_{A}\left( t\right) \geq \hat{V}_{A}\left( t\right) =V_{A}^{B}\left(
t\right) .
\end{equation*}%
Thus a risk-free liquidation increases the value of a 'derivative bond' to
the bond holder compared to the value that a bond has in the market
practice. Symmetrically, the value is reduced to the bond issuer, and this
reduction is an increasing function of the default risk of the bond holder.
This confirms that a substitution closeout guarantees the borrower, the
party which has payment to do in the future, making its risk of default the
only relevant variable that determines the value of a bond or loan. This is
a crucial feature for the stability of a debt market, but it has
implications that can appear paradoxical. An example is in the following.

\subsubsection{A special case with comonotonic defaults\label%
{section_comonotonicdefaults}}

Let us consider again the case of a risk free zero coupon bond, and assume
the total default dependence case, as represented by two co-monotonic
default times. We assume co-monotonicity as
\begin{equation*}
\tau ^{A}=\psi (\tau ^{B})
\end{equation*}%
for a deterministic and strictly increasing function $\psi $. Let us further
limit ourselves to situations where the lender is riskier than the borrower
in terms of probability of default. This is represented by assuming that $%
\psi (x)<x$, so that $\tau ^{A}<\tau ^{B}$ in all scenarios. This is clearly
a very extreme case; some example may happen if $B$ is somehow a subsidiary
of $A$, although in a real-world case a default time will never be exactly a
deterministic function of another default time (unless we consider the case
of simultaneous defaults, that in fact is what usually holds for
subsidiaries). Some models based on a physical explanation of default
dependency, such as structural models or the multivariate exponential
Marshall Olkin (1967) model we use in the following, do not even admit
comonotonicity outside the case of simultaneous default. However this is a
scenario admitted by other common models, like a bivariate default intensity
model where default intensities of $A$ and $B$ satisfy $\lambda ^{A}>\lambda
^{B}$ and where the two default time exponential triggers are connected by
the co-monotonic copula, or equivalently by a Gaussian copula with
correlation 1. Here default of a company triggers the default of a second
one in a fully predictable way, but the default of the second one may happen
any number of years after the default of the first one. The authors of this
paper have pointed out in Brigo and Chourdakis (2008) and in Morini (2009)
the risks of using models that admit comonotonic but not simultaneous
default times, a feature with scarce financial meaning that can lead to
misleading results. Here however we do not use this example because we think
it is realistic, but to point out some extreme consequences of the choice
about the closeout.

In the above setup we have that the risk free closeout formula (\ref%
{eq:VARFalt}) yields
\begin{equation*}
V_{A}(t)=e^{-\int\nolimits_{t}^{T}r\left( s\right) ds}
\end{equation*}%
whereas the substitution closeout formula yields
\begin{equation*}
\hat{V}_{A}(t)=e^{-\int\nolimits_{t}^{T}r\left( s\right) ds}\Pr_t (\tau
^{B}>T).
\end{equation*}%
We are in a situation where whenever there is default, $A$ defaults always
first. As a consequence, $A$ will never be impacted by $B$'s default, and
one could expect then the price of the bond to $A$ not to depend on the
default risk of $B$. While this happens with the risk-free closeout, this
does not happen with the substitution closeout, that maintains dependence on
default risk of $B$. We have to say, however, that although in this stylized
example it would make sense for the bond price not to depend on the default
probability of $B$, this does not happen in the market, where the bond price
remains $\hat{V}_{A}(t)$ regardless of default dependence issues, making the
price of bonds the same for all buyers.

\subsection{A simple two-name credit model}

To quantify the size of the above difference, and to analyze numerically the
practical effect of either assumption on closeout, we need to have a model
for the default times of our two names. Consistently with the purpose of
keeping complexity as low as possible since we are dealing with very
fundamental issues, we will use the simplest bivariate extension of the
standard single name credit model. Like in the single name market credit
model, we assume that the default time of the name $X$, $X\in \left\{
A,B\right\} $, is exponentially distributed. We take a flat default
intensity $\lambda _{X}$, so that the survival probability is%
\begin{equation*}
\Pr \left( \tau _{X}>T\right) =e^{-\lambda _{X}T},
\end{equation*}%
and for consistency also $r\left( s\right) $ is taken flat so that the
default-free bond is
\begin{equation*}
P_{T}=e^{-rT}.
\end{equation*}%
As baseline hypotheses, we take the two extreme scenarios, namely one
scenario where the variables $\tau ^{A}$ and $\tau ^{B}$ are independent,
and a second one where they are co-monotonic. We start with the independent
case. It is well known that in this case $\min \left( \tau ^{A},\tau
^{B}\right) $ is also exponentially distributed of parameter $\lambda
_{A}+\lambda _{B}$, since%
\begin{eqnarray*}
\Pr \left( \min \left( \tau ^{A},\tau ^{B}\right) >T\right) &=&\Pr \left(
\tau ^{A}>T\right) \Pr \left( \tau ^{B}>T\right) \\
&=&e^{-\left( \lambda _{A}+\lambda _{B}\right) T}
\end{eqnarray*}%
It is easy to compute the terms needed to apply Formula~(\ref{eq:VARFmain}).
First, one has

\begin{equation*}
\Pr \left( \tau ^{A}<\min(\tau ^{B},T)\right) =\frac{\lambda _{A}}{\lambda
_{A}+\lambda _{B}}\left( 1-e^{-\left( \lambda _{A}+\lambda _{B}\right)
T}\right),
\end{equation*}%
as one can easily show by solving the integral%
\begin{equation*}
\Pr \left( \{\tau ^{B} > \tau^{A}\} \cap \{T > \tau^{A}\}\right) =
\int_{0}^{T} \Pr(\tau^B > t) \Pr(\tau^A \in dt) = \int_{0}^{T}
e^{-\lambda_{B} t} \lambda_A e^{-\lambda_{A} t} dt
\end{equation*}%
We have also the special case%
\begin{eqnarray*}
\Pr \left( \tau ^{A}<\tau ^{B}\right) =\frac{\lambda _{A}}{\lambda
_{A}+\lambda _{B}}
\end{eqnarray*}%
which is obtained as a limit case of the earlier expression when $T \uparrow
\infty$. 
We have all we need for computing%
\begin{eqnarray}
&& V_{A}\left( t\right) =e^{-rT}\left\{ \Pr \left[ T<\min \left( \tau
^{A},\tau ^{B}\right) \right] +\Pr \left[ \tau ^{A}<\min \left( T,\tau
^{B}\right) \right] +R^{B}\Pr \left[ \tau ^{B}<\min \left( \tau
^{A},T\right) \right] \right\} =  \notag
\label{formula_biilateral_counterparty_for bond} \\
&& \hspace{1cm} = e^{-rT}\left\{ e^{-\left( \lambda _{A}+\lambda _{B}\right)
T}+\frac{\lambda _{A}}{\lambda _{A}+\lambda _{B}}\left( 1-e^{-\left( \lambda
_{A}+\lambda _{B}\right) T}\right) +R^{B}\frac{\lambda _{B}}{\lambda
_{A}+\lambda _{B}}\left( 1-e^{-\left( \lambda _{A}+\lambda _{B}\right)
T}\right) \right\}
\end{eqnarray}%
Notice that in our model the probability of events such as $\tau ^{A}=T$ or $%
\tau ^{A}=\tau ^{B}$ is zero.

In the comonotonic default case, we assume that
\begin{equation*}
\tau ^{A}=\xi /\lambda _{A},\ \ \tau ^{B}=\xi /\lambda _{B},
\end{equation*}%
with $\xi $ standard exponential random variable. It follows that
\begin{equation}
\tau ^{A}=\frac{\lambda _{B}}{\lambda _{A}}\tau ^{B}
\label{formula_comonotonicdefaults}
\end{equation}%
In this case it will be easy to compute the terms appearing in (\ref%
{eq:VARFmain}). We begin with the case where $\lambda _{B}>\lambda _{A}$.
Now $\tau ^{B}$ always happens first, so we have

\begin{equation}
V_{A}(t)=e^{-\int\nolimits_{t}^{T}r\left( s\right) ds}\left( \Pr_{t}[\tau
^{B}>T]+R^{B}\Pr_{t}[\tau ^{B}<T]\right) =\hat{V}_{A}(t).
\end{equation}%
Hence in this case risk free closeout and substitution closeout agree.
Furthermore, here it does not make sense to ask what happens at the default
time of the lender, because the borrower defaults first and closeout will
always happen before the default of the lender.

The interesting case where the two formulations disagree also in the
co-monotonic case is when $\lambda _{B}<\lambda _{A}$, and hence $\tau
^{B}>\tau ^{A}$. In this case one can see by looking at (\ref{eq:VARFalt})
and (\ref{formula_bilateral righr_for bond}) that%
\begin{equation*}
V_{A}(t)=e^{-\int\nolimits_{t}^{T}r\left( s\right) ds},
\end{equation*}%
which makes sense, given that default of $B$ will never happen to a solvent $%
A$, while the substitution closeout formula remains (\ref{formula_bilateral
righr_for bond}),
\begin{equation*}
\hat{V}_{A}(t)=e^{-\int\nolimits_{t}^{T}r\left( s\right) ds}\left[
\Pr_{t}\left( \tau ^{B}>T\right) +R^{B}\Pr_{t}\left( \tau ^{B}\leq T\right) %
\right] ,
\end{equation*}%
although we have to take into account Remark \ref{filtrationdelirium} if $%
t>0 $.

\subsection{What happens in case of default. An empirical example.}

\subsubsection{The independence case}

With this simple model we can test numerically the behaviour of the formula (%
\ref{formula_biilateral_counterparty_long}) or (\ref%
{formula_biilateral_counterparty_for bond}) with risk-free closeout. Set the
risk-free rate at $r=3\%$, and consider a bond with maturity $5$ years. The
price of the bond varies with the default risk of the borrower, as usual,
and here also with the default risk of the lender, due to the risk-free
closeout. In Figure 1 we show the price of the bond for intensities $\lambda
_{Lender}$, $\lambda _{Borrower}$ going from zero to 100\%. We consider $%
R^{Borrower}=0$ so that the level of the intensity approximately coincides
with the market CDS spread on the $5$ year maturity.

\bigskip

\begin{tabular}{l}
\includegraphics[width=8cm,height=13cm,angle=-90]{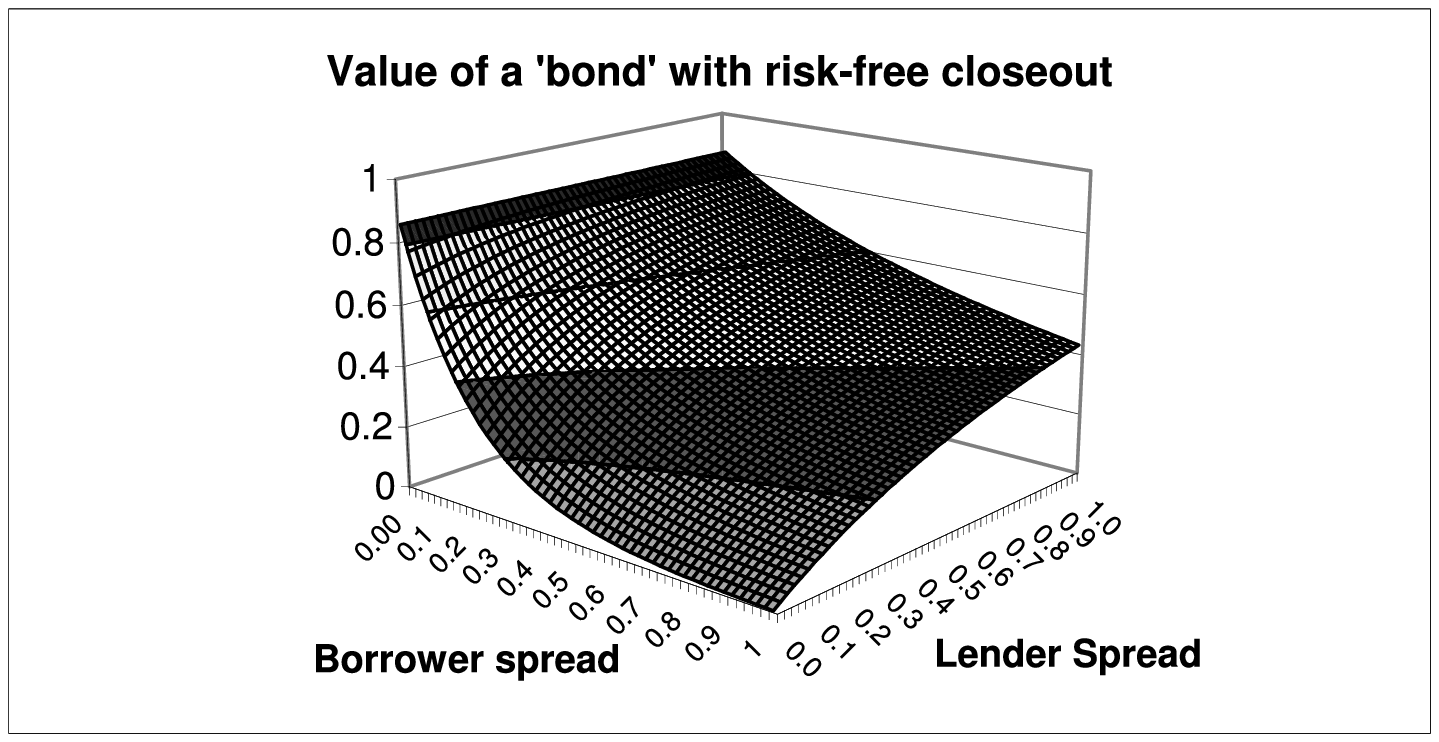} \\
Figure 1
\end{tabular}

\bigskip

\noindent We see that the effect of the lender's risk of default is not
negligible, and is particularly decisive when the borrower's risk is high.
Market operators should be aware of this consequence of a risk-free closeout.

The results of Figure 1 can be compared with those of Figure 2, where we
apply formula (\ref{formula_bilateral righr}) or (\ref{formula_bilateral
righr_for bond}) that assumes a substitution closeout.\bigskip

\begin{tabular}{l}
\includegraphics[width=7cm,height=13cm,angle=-90]{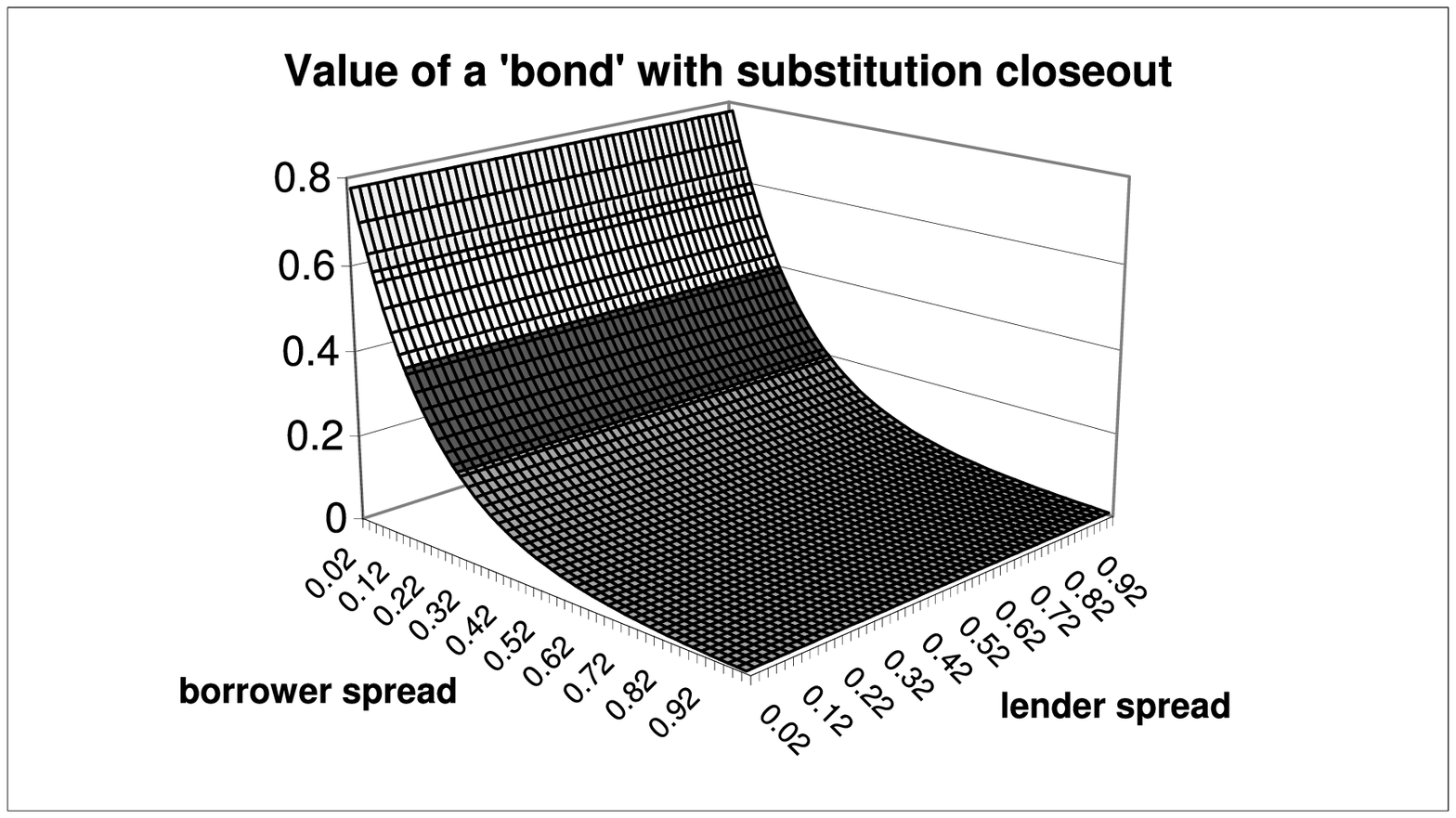} \\
Figure 2
\end{tabular}%
\bigskip

\noindent The pattern of Figure 2 traces precisely the pattern one would
expect from standard financial principles: independence of the price of the
deal from the risk of default of the counterparty which has no future
obligations in the deal.

There is a further implication of a risk-free closeout that is at odds with
market standards, and that should be pondered carefully by market operators
and regulators. When bilateral counterparty risk adjustment is computed as
indicated in the standard formula (\ref{formula_biilateral_counterparty_long}%
), that implies a risk-free closeout, we have a bizarre effect: \emph{a
company which has a net creditor position in a derivative and suddenly
defaults, can have a strong financial gain stemming from default itself},
whose benefit will go to its liquidators.

Symmetrically, and this is the most worrying part, the counterparty that
\emph{did not default and that is a net debtor of the defaulted company will
have to book a sudden loss due to default of the creditor}. Notice that this
is at odds with standard financial wisdom. It is natural that a \emph{%
creditor} of company is damaged by the default of the company, but here we
have something more: when (\ref{formula_biilateral_counterparty_long}) is
used, there is a damage not only to creditors, but also to \emph{debtors}.
This is worrying since it implies that with a risk-free closeout the default
contagion spreads also to net debtors, not only to net creditors.

Let us observe a numerical example. We start from the above $r=3\%$ and a
bond with maturity $5$ years, for a $1$bn notional. Now we take $%
R^{Borrower}=20\%$ and two risky parties. We suppose the borrower has a very
low credit quality, as expressed by $\lambda _{Borrower}=0.2$, that means
\begin{equation*}
\Pr \left( \tau ^{Borrower}\leq 5y\right) =63.2\%.
\end{equation*}%
The lender has a much higher credit quality, as expressed by $\lambda
_{Lender}=0.04$. This means a lower probability of default, that however is
not negligible, being%
\begin{equation*}
\Pr \left( \tau ^{Lender}\leq 5y\right) =18.1\%\text{.}
\end{equation*}%
A risk free bond issuance with the same maturity and notional would cost
\begin{equation*}
P_{T}=860.7mn
\end{equation*}%
Using the formula with risk-free closeout, we get that a risky bond, within
the two parties above, has price%
\begin{equation*}
V_{Lender}=359.5mn
\end{equation*}%
to be compared with the price coming form the formula with substitution
closeout%
\begin{equation*}
\hat{V}_{Lender}=316.6mn.
\end{equation*}%
These figures can be very easily computed via (\ref%
{formula_biilateral_counterparty_long}) or more directly with (\ref%
{formula_biilateral_counterparty_for bond}). The example confirms a
difference in the valuation given by the two formulas. The higher value of $%
V_{Lender}$ depends on the probability of default of the lender.

The difference of the two valuations is not negligible but not dramatic.
More relevant is the difference of what happens in case of a default under
the two assumptions on closeout. We have the following risk-adjusted
probabilities on the happening of a default event%
\begin{eqnarray*}
\Pr \left[ \tau ^{Borrower}<\min \left( 5y,\tau ^{Lender}\right) \right]
&=&58\%, \\
\Pr \left[ 5y<\min \left( \tau ^{Lender},\tau ^{Borrower}\right) \right]
&=&30\%, \\
\Pr \left[ \tau ^{Lender}<\min \left( 5y,\tau ^{Borrower}\right) \right]
&=&12\%.
\end{eqnarray*}%
The two formulas agree on what happens in case of no default or in case of
default of the borrower first. These are the most likely scenarios,
totalling $88\%$ probability. But with a non-negligible probability, $12\%$,
the lender can default first. Let us analyze in detail what happens in this
case. Suppose the exact day when default happens is
\begin{equation*}
\tau ^{Lender}=2.5years.
\end{equation*}%
Just before default, at $2.5$ years less one day, we have for the borrower
the following book value of the liability produced by the above deal:%
\begin{equation*}
V_{Borrower}\left( \tau ^{Lender}-1d\right) =-578.9\text{mln}
\end{equation*}%
if he assumes a risk-free closeout, or%
\begin{equation*}
\hat{V}_{Borrower}\left( \tau ^{Lender}-1d\right) =-562.7\text{mln}
\end{equation*}%
if he assumes a substitution closeout.

Now default of the lender happens. In case of a risk-free closeout, one can
easily check in (\ref{formula_biilateral_counterparty_long}) or (\ref%
{formula_biilateral_counterparty_for bond}), calling the lender $A$ and $B$
the borrower, that when $\tau ^{A}\leq \min \left( T,\tau ^{B}\right) $ and
\begin{equation*}
V_{B}^{0}\left( \tau ^{A}\right) \leq 0\text{,}
\end{equation*}%
like here, we have%
\begin{equation*}
V_{B}\left( \tau ^{A}\right) =V_{B}^{0}\left( \tau ^{A}\right) \text{.}
\end{equation*}%
Thus the book value of the bond becomes simply the value of a risk free bond,%
\begin{equation*}
V_{B}\left( \tau ^{A}+1d\right) =-927.7\text{mln.}
\end{equation*}%
The borrower, which has not defaulted, must pay this amount entirely to the
defaulted lender. He has a sudden loss of
\begin{eqnarray*}
&&927.7\text{mln}-578.9\text{mln} \\
&=&348.8\text{mln}
\end{eqnarray*}%
due to default of the lender. This way of regulating bilateral counterparty
risk leads to a default contagion that extends not only to the creditors,
but also, and in a more sudden way, to the debtors of the defaulted company.
This is a worrying effect that regulators can hardly consider desirable.
Additionally, formulas (\ref{formula_biilateral_counterparty_long}) and (\ref%
{formula_biilateral_counterparty_for bond}) show clearly that the loss is
higher the higher the credit spreads, and this is what makes this kind of
contagion particularly dangerous. Finally, notice that this book loss of $%
365 $mln will soon become a material, realized loss because the borrower has
to pay it to the liquidators of the defaulted lender, well before the
original maturity of the deal.

Furthermore, this behaviour is not consistent with financial practice in the
bond and loan market, where there are no big losses to the borrower or bond
issuer due to the default of a lender or a bond-holder. This makes the
behaviour particularly undesirable. We mentioned at the beginning of this
section that with bilateral counterparty risk banks can book a profit when
their own risk of default increases, and that this somewhat disturbing fact
has been accepted also because it is consistent with the fair value option
in the booking of bonds. Here we have found another disturbing effect when a
risk-free closeout is considered: a sudden and big gain for the lender's
liquidators when the lender himself defaults, and a symmetric big loss for
the borrower. This is even more disturbing, and it is not even consistent
with market practice on bonds.

The bond practice is instead consistent with the substitution closeout. In
this case, applying (\ref{formula_bilateral righr}) we have%
\begin{equation*}
\hat{V}_{Borrower}\left( \tau ^{Lender}+1d\right) =-578.9\text{mln.}
\end{equation*}%
There is no discontinuity and no loss for the borrower in case of default of
the lender.

We point out that the above example could be extended easily to those more
complex derivative payoffs where there are, like in the 'derivative' bond, a
clear lender and a clear borrower that do not change through the life of the
deal. A party $X$ is a lender when $V_{X}^{0}\left( t\right) >0$, $T>t>0$,
while he is a borrower if $V_{X}^{0}\left( t\right) <0$, $T>t>0$. Payoffs
that generate this situation are very common. The simplest example is an
option contract. The option buyer is the lender, the option writer is the
borrower.

Additionally, some time after inception all deals, even the most complex
derivatives, can have mark-to-market far away from zero, so that for one
party $V_{X}^{0}\left( t\right) \ll 0$. Such party becomes a net borrower
and in case of default of the other party would suffer, following a
risk-free closeout, losses similar to those outlined above.

\subsubsection{The co-monotonic case}

We now take an example where the two default times have constant intensities
and are co-monotonic, and we will see, conversely, that this time the
substitution closeout has a dramatic effect for the creditors and
liquidators of the defaulted company.

We assume $\lambda _{Borrower}=3.6\%$, $\lambda _{Lender}=4\%$, $r=3\%$, $%
T=5y$ and $R_{Borrower}=0$. This setting means that $Lender$ always defaults
first, since following formula (\ref{formula_comonotonicdefaults})%
\begin{equation*}
\tau ^{Borrower}=\frac{\lambda _{Lender}}{\lambda _{Borrower}}\tau
^{Lender}=1.11\tau ^{Lender}
\end{equation*}
We still assume a notional of 1bn. The initial value to the borrower of the
bond under bilateral counterparty risk for the risk free closeout is%
\begin{equation*}
-e^{-0.03\cdot 5}1bn=-860.71mn,
\end{equation*}%
namely the risk-free value, consistently with the example we showed in
Section \ref{section_comonotonicdefaults}. The value of the bond under
substitution closeout is%
\begin{equation*}
-e^{-(0.066)\cdot 5}1bn=-718.92mn
\end{equation*}%
Now let us see what happens when the lender A defaults at 2.5y. We need to
remember that we are in a situation where A defaults always before B, and
where we know exactly that, if A defaults at 2.5y, B defaults in 2.77y. This
is clearly a purely toy case, but it approximates a possibly realistic one:
the case of a company B that, as an effect of the default of A, sees its own
risk of default increase dramatically, going to the verge of default.

In such a case the risk free closeout gives us the same value both
immediately before and after default, namely, to the lender A:%
\begin{equation*}
e^{-0.03\cdot 2.5}1bn=927.74mn
\end{equation*}%
with no discontinuity, whereas the substitution closeout implies a jump from%
\begin{eqnarray*}
&&e^{-0.03\cdot 2.5}\Pr \left( \tau ^{Borrower}>5y|\tau
^{Lender}>2.5y\right) 1bn \\
&=&e^{-0.03\cdot 2.5}\Pr \left( \tau ^{Lender}>\frac{5}{1.1}y|\tau
^{Lender}>2.5y\right) 1bn \\
&=&e^{-0.03\cdot 2.5}e^{-0.04\left( \frac{5}{1.\bar{1}}-2.5\right) }=856.41bn
\end{eqnarray*}%
to
\begin{equation*}
e^{-0.03\cdot 2.5}\Pr \left( \tau ^{Borrower}>5y|\tau ^{Lender}=2.5y\right)
1bn=0.
\end{equation*}%
Due to the fact that the model implies that the borrower will default at
2.77y, which is before the maturity, the substitution closeout, that takes
into account the default risk of the counterparty, sees a null value for the
bond.

In other terms, at the default of the lender, the creditors and liquidators
of the lender see all their claims towards the borrower lose their value.
This confirms that the substitution closeout reduces the claims of the
creditors towards the debtors of the defaulted entity, and this effect is
stronger the stronger is the default dependence between the defaulted entity
and its borrowers. This may be the case, for example, when the defaulted
entity is a company with a relevant systemic impact.

\section{The collateral\label{section_collateral}}

There is one final thing to assess, and it is the behaviour of the two forms
of closeout for collateralized deals. A defaultable deal should satisfy the
following property: the combination of the defaultable deal and a collateral
agreement must eliminate losses at default.

In a standardized collateral agrement for a deal with payoff $X$ the
counterparty which is a net borrower provides the net lender with an amount
of liquidity equal to the risk-free net present value of $X$. This amount of
collateral generates interest and is regularly updated to remain equal to
the net present value of $X$, and can be claimed back by the borrower only
at maturity of the deal if the borrower has not defaulted earlier. If
instead the borrower defaults earlier, the collateral must be used to offset
the default loss.

We analyze in detail what happens at default considering a very simple
payoff where $X=1$ and must be paid at $T$, namely the simple 'derivative'
bond considered above. The borrower must give at time $0$ to the lender an
amount of liquidity equal to the risk-free net present value of the future
payoff,
\begin{equation*}
C\left( 0\right) =e^{-\int\nolimits_{0}^{T}r\left( s\right) ds}.
\end{equation*}%
Then the borrower must update regularly this quantity, in such a way that
the collateral remains equal at any time $t$ to the net present value of the
future payoff,
\begin{equation*}
C\left( t\right) =e^{-\int\nolimits_{t}^{T}r\left( s\right) ds}.
\end{equation*}%
The collateral is updated daily. We approximate this daily settlement with a
continuous settlement. In this case, in order to keep the collateral $%
C\left( t\right) $ at $e^{-\int\nolimits_{t}^{T}r\left( s\right) ds}$, the
borrower must pay to the lender continuously an amount%
\begin{equation}
r\left( t\right) C\left( t\right) dt,  \label{formula_settlement payment}
\end{equation}%
in fact it is trivial to show that if%
\begin{eqnarray*}
dC\left( t\right) &=&r\left( t\right) C\left( t\right) dt, \\
C\left( 0\right) &=&e^{-\int\nolimits_{0}^{T}r\left( s\right) ds},
\end{eqnarray*}%
then%
\begin{equation*}
C\left( t\right) =e^{-\int\nolimits_{t}^{T}r\left( s\right) ds},
\end{equation*}%
as desired. On the other hand, the lender that keeps the collateral $C\left(
t\right) $ must give back to the borrower the interest generated
continuously by $C\left( t\right) $. This transforms into a continuous
payment%
\begin{equation}
r\left( t\right) C\left( t\right) dt.  \label{formula_interespt payment}
\end{equation}%
made by the lender to the borrower.

We immediately notice that the settlement payment (\ref{formula_settlement
payment}) made by the borrower cancels out with the interest payment (\ref%
{formula_interespt payment}) made by the lender. Thus for a simple payoff
the collateral agreement becomes a sort of self-financing strategy, namely
it does not involve any net continuous exchange of money. The borrower $B$
pays $e^{-\int\nolimits_{0}^{T}r\left( s\right) ds}$ at time $0$. The
collateral remains with the lender and earns interest in time at rate $%
r\left( t\right) $, this interest $r\left( t\right) C\left( t\right) dt$ is
not returned to the borrower but added to the value of the collateral, that
in this way is always
\begin{equation*}
C\left( t\right) =e^{-\int\nolimits_{t}^{T}r\left( s\right) ds}.
\end{equation*}

\subsection{Risk-free closeout and collateral}

If we consider the risk-free closeout, the requirement of a match between
the closeout and the value of the collateral at the first default $\tau $
among those of the two counterparties appears trivially obtained. The
risk-free closeout amount is always
\begin{equation*}
V_{A}\left( \tau \right) =e^{-\int\nolimits_{\tau }^{T}r\left( s\right) ds}.
\end{equation*}
The value of the collateral at $\tau $ is always%
\begin{equation*}
C\left( \tau \right) =e^{-\int\nolimits_{\tau }^{T}r\left( s\right) ds}.
\end{equation*}%
If we just say that the collateral is an amount of money to be used to
offset the default closeout, it turns out that the lender always has the
right amount of money to offset the default closeout.

\subsection{Substitution closeout and collateral}

However, if we take a less naive view we remember that collateral is not
just an amount of money but a contract with specific features. This contract
has a value at default and it is this value, not the amount of money held by
the lender, that needs to match the closeout amount. If this happens, there
will be no transfer on money at default of a collateralized party and no
losses for any party, exactly as we expect for a collateralized deal. We
show in the following that this can be obtained rather easily with a
substitution closeout . If we think of the practical working of collateral,
we can treat collateral as an amount of money given to the lender from the
borrower coupled with a claim of the borrower towards the lender to have
this collateral back. If there is no default of the borrower before
maturity, the borrower has a claim to have the collateral back at
termination of the deal, namely a payoff $C\left( T\right) $ at maturity $T$
upon no default:%
\begin{equation*}
1_{\left\{ \tau ^{B}>T\right\} }C\left( T\right) =1_{\left\{ \tau
^{B}>T\right\} }\ \ \ \ \ \ \ \ \text{at }T.
\end{equation*}%
If instead the default of the borrower happens before maturity, the borrower
has a claim to receive the collateral back at its own default time $\tau
^{B} $
\begin{equation*}
1_{\left\{ \tau ^{B}<T\right\} }C\left( \tau \right) =1_{\left\{ \tau
^{B}<T\right\} }e^{-\int\nolimits_{\tau ^{B}}^{T}r\left( s\right) ds}\ \ \ \
\ \ \ \ \text{at }\tau ^{B}.
\end{equation*}%
In this simple setting we can easily check what happens when there is a
default, either of the borrower or of the lender. In case the borrower
defaults first, the closeout amount that the borrower should give to the
lender is
\begin{equation*}
e^{-\int\nolimits_{\tau ^{B}}^{T}r\left( s\right) ds}
\end{equation*}%
At the same time the borrower has the right to receive back the collateral,
which has a value%
\begin{equation*}
e^{-\int\nolimits_{\tau ^{B}}^{T}r\left( s\right) ds}
\end{equation*}%
There is a perfect netting and a perfect match of collateral value and
closeout amount of the deal at default. No money is transferred and no loss
is suffered, as it should be for collateralized deals. This holds no matter
whether we consider risk-free or substitution closeout, since we are
considering the default of the borrower. At default of the borrower the
residual net present value is risk free under either closeout.

More interesting is the case of the default of the lender at $\tau ^{L}\leq
T $, $\tau ^{B}$. The borrower has a claim to receive back the collateral at
maturity if the borrower itself does not default. The discounted expected
value of such a claim is%
\begin{equation}
\mathbb{E}_{\tau ^{L}}\left[ e^{-\int\nolimits_{\tau ^{L}}^{T}r\left(
s\right) ds}1_{\left\{ \tau ^{B}>T\right\} }\right] .
\label{formula_discounted_collateral}
\end{equation}%
The claim of the lender depends on the choice about the closeout amount. In
case of substitution closeout, the value of the lender's claim is the
discounted expected payoff of the deal taking into account the possibility
that $\tau ^{B}$ defaults in the future,%
\begin{equation*}
\mathbb{E}_{\tau ^{L}}\left[ e^{-\int\nolimits_{\tau ^{L}}^{T}r\left(
s\right) ds}1_{\left\{ \tau ^{B}>T\right\} }\right] .
\end{equation*}%
Again there is a perfect netting and a perfect match of collateral value and
closeout amount of the deal at default. No money is transferred and no loss
is suffered, as it should be for collateralized deals. When no default
happens before maturity, the lender gives the collateral back to the
borrower, again consistently with the practical working of collateral.

\section{Conclusions}

In this paper we have analyzed the effect of the assumptions about the
computation of the closeout amount on the counterparty risk adjustments of
derivatives. We have compared the risk-free closeout assumed in the earlier
literature with the substitution closeout we introduce here, which is
inspired by the financial rationale contained in the recent ISDA
documentation on the subject.

We have provided a formula for bilateral counterparty risk when a
substitution closeout is used at default. We have reckoned that the
substitution closeout is consistent with counterparty risk adjustments for
standard and consolidated financial products such as bonds and loans. On the
contrary the risk-free closeout introduces at time $0$ a dependence on the
risk of default of the party with no future obligations.

We have also shown that in case of risk-free closeout a party that is a net
debtor of a company will have a sudden loss at the default of the latter,
and this loss is higher the higher the debtor's credit spreads. This does
not happen when a substitution closeout is considered.

Thus, the risk-free closeout increases the number of operators subject to
contagion from a default, including parties that currently seem not to think
they are exposed, and this is certainly a negative fact. On the other hand,
it spreads the default losses on higher number of parties and reduces the
classic contagion channel affecting creditors. For the creditors, this is a
positive fact because it brings more money to the liquidators of the
defaulted company. Additionally, a risk-free closeout is easier to implement
at default time, since it corresponds to the net present value of a
collateralized version of the deal. Thanks to this, one can also find it
easier to understand why the provision of collateral makes a defaultable
deal to become default-free. In the end, however, we have shown why if we
take into account the contractual features of a collateral agreement, the
provision of collateral makes a deal default free also when substitution
closeout applies.

We think that the closeout issue should be considered carefully by market
operators and regulators. For example, if the risk-free closeout introduced
in the previous literature had to be recognized as a standard, banks should
understand the consequences of this. In fact banks usually perform stress
tests and set aside reserves for the risk of default of their net borrowers,
but do not consider any risk related to the default of net lenders. The
above computations, and the numerical examples, show that, should a
risk-free closeout prevail, banks had better set aside important reserves
against this risk. On the other hand, under substitution closeout, banks can
expect the recovery to be lowered when their net borrowers default, compared
to the case when a risk free closeout applies. In case of a substitution
closeout, in fact, the money collected by liquidators from the
counterparties will be lower, since deflated by the default probability of
the counterparties themselves, and this will be even more evident if they
are strongly correlated to the defaulted entity.

\section{Appendix}

First of all, notice that from formula (\ref{formula_bilateral righr}) one
can compute that%
\begin{eqnarray*}
\hat{V}_{B}\left( t\right) &=&\mathbb{E}_{t}\left\{ 1_{0}\Pi _{B}\left(
t,T\right) \right\} \\
&&+\mathbb{E}_{t}\left\{ 1_{B}\left[ \Pi _{B}\left( t,\tau ^{B}\right)
+D\left( t,\tau ^{B}\right) \left( \left( V_{B}^{A}\left( \tau ^{B}\right)
\right) ^{+}-R^{B}\left( -V_{B}^{A}\left( \tau ^{B}\right) \right)
^{+}\right) \right] \right\} \\
&&+\mathbb{E}_{t}\left\{ 1_{A}\left[ \Pi _{B}\left( t,\tau ^{A}\right)
+D\left( t,\tau ^{A}\right) \left( R^{A}\left( V_{B}^{B}\left( \tau
^{A}\right) \right) ^{+}-\left( -V_{B}^{B}\left( \tau ^{A}\right) \right)
^{+}\right) \right] \right\} . \\
&=&-\mathbb{E}_{t}\left\{ 1_{0}\Pi _{A}\left( t,T\right) \right\} \\
&&+\mathbb{E}_{t}\left\{ 1_{B}\left[ -\Pi _{A}\left( t,\tau ^{B}\right)
+D\left( t,\tau ^{B}\right) \left( \left( -V_{A}^{A}\left( \tau ^{B}\right)
\right) ^{+}-R^{B}\left( V_{A}^{A}\left( \tau ^{B}\right) \right)
^{+}\right) \right] \right\} \\
&&+\mathbb{E}_{t}\left\{ 1_{A}\left[ -\Pi _{A}\left( t,\tau ^{A}\right)
+D\left( t,\tau ^{A}\right) \left( R^{A}\left( -V_{A}^{B}\left( \tau
^{A}\right) \right) ^{+}-\left( V_{A}^{B}\left( \tau ^{A}\right) \right)
^{+}\right) \right] \right\} \\
&=&-\hat{V}_{A}\left( t\right) ,
\end{eqnarray*}%
so we have the desired symmetry property, like in the standard formula (\ref%
{formula_biilateral_counterparty_long}).

The standard formula (\ref{formula_biilateral_counterparty_long}) with
risk-free closeout can be simplified to
\begin{eqnarray}
&&\hat{V}_{A}\left( t\right)
\begin{tabular}{l}
$=$%
\end{tabular}%
V_{A}^{0}\left( t\right)  \label{formula_biilateral_counterparty} \\
&&+\mathbb{E}_{t}\left[ L^{A}1_{A}D\left( t,\tau ^{A}\right) \cdot \left(
-V_{A}^{0}\left( \tau ^{A}\right) \right) ^{+}\right]  \notag \\
&&-\mathbb{E}_{t}\left[ L^{B}1_{B}D\left( t,\tau ^{B}\right) \cdot \left(
V_{A}^{0}\left( \tau ^{B}\right) \right) ^{+}\right] .  \notag
\end{eqnarray}%
The alternative formula (\ref{formula_bilateral righr}) with
replacement-cost closeout does not allow such a trivial simplification.
However some simplifications are possible by replacing $V_{A}^{B}\left( \tau
^{1}\right) $ and $V_{A}^{A}\left( \tau ^{B}\right) $ with their
definitions, obtaining%
\begin{equation*}
\begin{tabular}{l}
$\hat{V}_{A}\left( t\right)
\begin{tabular}{l}
$=$%
\end{tabular}%
\mathbb{E}_{t}\left\{ 1_{0}\Pi _{A}\left( t,T\right) \right\} \medskip $ \\
$+\mathbb{E}_{t}\left\{ 1_{A}\left[ \Pi _{A}\left( t,\tau ^{A}\right)
+D\left( t,\tau ^{A}\right) \left( \left( V_{A}^{0}\left( \tau ^{A}\right) -%
\mathbb{E}_{\tau ^{A}}\left[ L^{B}1_{\left\{ \tau ^{B}\leq T\right\}
}D\left( \tau ^{A},\tau ^{B}\right) \cdot \left( V_{A}^{0}\left( \tau
^{B}\right) \right) ^{+}\right] \right) ^{+}\right. \right. \right. \medskip
$ \\
$\left. \left. \left. -R^{A}\left( -V_{A}^{0}\left( \tau ^{A}\right) +%
\mathbb{E}_{\tau ^{A}}\left[ L^{B}1_{\left\{ \tau ^{B}\leq T\right\}
}D\left( \tau ^{A},\tau ^{B}\right) \cdot \left( V_{A}^{0}\left( \tau
^{B}\right) \right) ^{+}\right] \right) ^{+}\right) \right] \right\}
\medskip $ \\
$+\mathbb{E}_{t}\left\{ 1_{B}\left[ \Pi _{A}\left( t,\tau ^{B}\right)
+D\left( t,\tau ^{B}\right) \left( R^{B}\left( V_{A}^{0}\left( \tau
^{B}\right) +\mathbb{E}_{\tau ^{B}}\left[ L^{A}1_{\left\{ \tau ^{A}\leq
T\right\} }D\left( \tau ^{B},\tau ^{A}\right) \cdot \left( -V_{A}^{0}\left(
\tau ^{A}\right) \right) ^{+}\right] \right) ^{+}\right. \right. \right.
\medskip $ \\
$\left. \left. \left. -\left( -V_{A}^{0}\left( \tau ^{B}\right) -\mathbb{E}%
_{\tau ^{B}}\left[ L^{A}1_{\left\{ \tau ^{A}\leq T\right\} }D\left( \tau
^{B},\tau ^{A}\right) \cdot \left( -V_{A}^{0}\left( \tau ^{A}\right) \right)
^{+}\right] \right) ^{+}\right) \right] \right\} .$%
\end{tabular}%
\end{equation*}%
If we replace
\begin{equation*}
1_{0}=1-1_{A}-1_{B},
\end{equation*}%
recalling%
\begin{equation*}
\Pi _{A}\left( t,\tau \right) -\Pi _{A}\left( t,T\right) =-D\left( t,\tau
\right) \Pi _{A}\left( \tau ,T\right)
\end{equation*}%
we obtain%
\begin{equation*}
\begin{tabular}{l}
$\hat{V}_{A}\left( t\right)
\begin{tabular}{l}
$=$%
\end{tabular}%
\mathbb{E}_{t}\left\{ \Pi _{A}\left( t,T\right) \right\} \medskip $ \\
$+\mathbb{E}_{t}\left\{ 1_{A}\left[ -D\left( t,\tau ^{A}\right) \Pi
_{A}\left( \tau ^{A},T\right) +D\left( t,\tau ^{A}\right) \left( \left(
V_{A}^{0}\left( \tau ^{A}\right) -\mathbb{E}_{\tau ^{A}}\left[
L^{B}1_{\left\{ \tau ^{B}\leq T\right\} }D\left( \tau ^{A},\tau ^{B}\right)
\cdot \left( V_{A}^{0}\left( \tau ^{B}\right) \right) ^{+}\right] \right)
^{+}\right. \right. \right. \medskip $ \\
$\left. \left. \left. -R^{A}\left( -V_{A}^{0}\left( \tau ^{A}\right) +%
\mathbb{E}_{\tau ^{A}}\left[ L^{B}1_{\left\{ \tau ^{B}\leq T\right\}
}D\left( \tau ^{A},\tau ^{B}\right) \cdot \left( V_{A}^{0}\left( \tau
^{B}\right) \right) ^{+}\right] \right) ^{+}\right) \right] \right\}
\medskip $ \\
$+\mathbb{E}_{t}\left\{ 1_{B}\left[ -D\left( t,\tau ^{B}\right) \Pi
_{A}\left( \tau ^{B},T\right) +D\left( t,\tau ^{B}\right) \left( R^{B}\left(
V_{A}^{0}\left( \tau ^{B}\right) +\mathbb{E}_{\tau ^{B}}\left[
L^{A}1_{\left\{ \tau ^{A}\leq T\right\} }D\left( \tau ^{B},\tau ^{A}\right)
\cdot \left( -V_{A}^{0}\left( \tau ^{A}\right) \right) ^{+}\right] \right)
^{+}\right. \right. \right. \medskip $ \\
$\left. \left. \left. -\left( -V_{A}^{0}\left( \tau ^{B}\right) -\mathbb{E}%
_{\tau ^{B}}\left[ L^{A}1_{\left\{ \tau ^{A}\leq T\right\} }D\left( \tau
^{B},\tau ^{A}\right) \cdot \left( -V_{A}^{0}\left( \tau ^{A}\right) \right)
^{+}\right] \right) ^{+}\right) \right] \right\} .$%
\end{tabular}%
\end{equation*}%
Now we use
\begin{equation*}
\left( X\right) ^{+}-R\left( -X\right) ^{+}=X+\left( 1-R\right) \left(
-X\right) ^{+},
\end{equation*}%
and%
\begin{equation*}
R\left( X\right) ^{+}-\left( -X\right) ^{+}=X-\left( 1-R\right) \left(
X\right) ^{+},
\end{equation*}%
getting%
\begin{equation*}
\begin{tabular}{l}
$\hat{V}_{A}\left( t\right)
\begin{tabular}{l}
$=$%
\end{tabular}%
\mathbb{E}_{t}\left\{ \Pi _{A}\left( t,T\right) \right\} \medskip $ \\
$+\mathbb{E}_{t}\left\{ 1_{A}\left[ -D\left( t,\tau ^{A}\right) \Pi
_{A}\left( \tau ^{A},T\right) +D\left( t,\tau ^{A}\right) \left(
V_{A}^{0}\left( \tau ^{A}\right) -\mathbb{E}_{\tau ^{A}}\left[
L^{B}1_{\left\{ \tau ^{B}\leq T\right\} }D\left( \tau ^{A},\tau ^{B}\right)
\cdot \left( V_{A}^{0}\left( \tau ^{B}\right) \right) ^{+}\right] \right.
\right. \right. \medskip $ \\
$\left. \left. \left. +\left( 1-R^{A}\right) \left( -V_{A}^{0}\left( \tau
^{A}\right) +\mathbb{E}_{\tau ^{A}}\left[ L^{B}1_{\left\{ \tau ^{B}\leq
T\right\} }D\left( \tau ^{A},\tau ^{B}\right) \cdot \left( V_{A}^{0}\left(
\tau ^{B}\right) \right) ^{+}\right] \right) ^{+}\right) \right] \right\}
\medskip $ \\
$+\mathbb{E}_{t}\left\{ 1_{2}\left[ -D\left( t,\tau ^{B}\right) \Pi
_{A}\left( \tau ^{B},T\right) +D\left( t,\tau ^{B}\right) \left(
V_{A}^{0}\left( \tau ^{B}\right) +\mathbb{E}_{\tau ^{B}}\left[
L^{A}1_{\left\{ \tau ^{A}\leq T\right\} }D\left( \tau ^{B},\tau ^{A}\right)
\cdot \left( -V_{A}^{0}\left( \tau ^{A}\right) \right) ^{+}\right] \right.
\right. \right. \medskip $ \\
$\left. \left. \left. -\left( 1-R^{B}\right) \left( V_{A}^{0}\left( \tau
^{B}\right) +\mathbb{E}_{\tau ^{B}}\left[ L^{A}1_{\left\{ \tau ^{A}\leq
T\right\} }D\left( \tau ^{B},\tau ^{A}\right) \cdot \left( -V_{A}^{0}\left(
\tau ^{A}\right) \right) ^{+}\right] \right) ^{+}\right) \right] \right\} .$%
\end{tabular}%
\end{equation*}%
Now we apply the law of iterated expectation, exploiting%
\begin{eqnarray*}
&&\mathbb{E}_{t}\left[ -D\left( t,\tau \right) \Pi _{A}\left( \tau ,T\right)
+D\left( t,\tau \right) V_{A}^{0}\left( \tau \right) \right] \\
&=&\mathbb{E}_{t}\left[ -D\left( t,\tau \right) \Pi _{A}\left( \tau
,T\right) +D\left( t,\tau \right) \mathbb{E}_{\tau }\left[ \Pi _{A}\left(
\tau ,T\right) \right] \right] \\
&=&\mathbb{E}_{t}\left[ -D\left( t,\tau \right) \mathbb{E}_{\tau }\left[ \Pi
_{A}\left( \tau ,T\right) \right] +D\left( t,\tau \right) \mathbb{E}_{\tau }%
\left[ \Pi _{A}\left( \tau ,T\right) \right] \right] =0,
\end{eqnarray*}%
so that%
\begin{equation*}
\begin{tabular}{l}
$\hat{V}_{A}\left( t\right)
\begin{tabular}{l}
$=$%
\end{tabular}%
\mathbb{E}_{t}\left\{ \Pi _{A}\left( t,T\right) \right\} \medskip $ \\
$+\mathbb{E}_{t}\left\{ 1_{1}\left[ D\left( t,\tau ^{A}\right) \left( -%
\mathbb{E}_{\tau ^{A}}\left[ L^{B}1_{\left\{ \tau ^{B}\leq T\right\}
}D\left( \tau ^{A},\tau ^{B}\right) \cdot \left( V_{A}^{0}\left( \tau
^{B}\right) \right) ^{+}\right] \right. \right. \right. \medskip $ \\
$\left. \left. \left. +\left( 1-R^{A}\right) \left( -V_{A}^{0}\left( \tau
^{A}\right) +\mathbb{E}_{\tau ^{A}}\left[ L^{B}1_{\left\{ \tau ^{B}\leq
T\right\} }D\left( \tau ^{A},\tau ^{B}\right) \cdot \left( V_{A}^{0}\left(
\tau ^{B}\right) \right) ^{+}\right] \right) ^{+}\right) \right] \right\}
\medskip $ \\
$+\mathbb{E}_{t}\left\{ 1_{2}\left[ D\left( t,\tau ^{B}\right) \left(
\mathbb{E}_{\tau ^{B}}\left[ L^{A}1_{\left\{ \tau ^{A}\leq T\right\}
}D\left( \tau ^{B},\tau ^{A}\right) \cdot \left( -V_{A}^{0}\left( \tau
^{A}\right) \right) ^{+}\right] \right. \right. \right. \medskip $ \\
$\left. \left. \left. -\left( 1-R^{B}\right) \left( V_{A}^{0}\left( \tau
^{B}\right) +\mathbb{E}_{\tau ^{B}}\left[ L^{A}1_{\left\{ \tau ^{A}\leq
T\right\} }D\left( \tau ^{B},\tau ^{A}\right) \cdot \left( -V_{A}^{0}\left(
\tau ^{A}\right) \right) ^{+}\right] \right) ^{+}\right) \right] \right\} .$%
\end{tabular}%
\end{equation*}

\end{document}